\documentclass[11pt]{article}

\newsavebox{\foobox}
\newcommand{\slantbox}[2][0]{\mbox{%
        \sbox{\foobox}{#2}%
        \hskip\wd\foobox
        \pdfsave
        \pdfsetmatrix{1 0 #1 1}%
        \llap{\usebox{\foobox}}%
        \pdfrestore
}}
\newcommand\unslant[2][-.25]{\slantbox[#1]{$#2$}}

\newcommand{\mmu}{\text{\unslant\mu}}
\newcommand{\mpi}{\text{\unslant[-.18]\pi}}
\newcommand{\mdelta}{\text{\unslant[-.18]\delta}}

\usepackage[left=2cm, right=2cm, top=2.5cm, bottom=2.5cm]{geometry}
\usepackage[x11names]{xcolor}
\usepackage{fancyhdr, amssymb, cancel, amsmath, graphicx, pgfplots, tikz}
\usepackage{isomath}

\usetikzlibrary{shadows}

\newcommand{\stylecolor}{blue!60!black}

\usepackage[labelfont={bf,sf, color=\stylecolor}, margin={1.5cm,0cm}]{caption}

\usepackage[colorlinks=true, urlcolor=\stylecolor!70!white, linkcolor=\stylecolor, citecolor=\stylecolor!70!white, hyperindex=true, linktocpage=true]{hyperref}

\usepackage{amsthm}
\newtheoremstyle{theor}{10pt}{10pt}{}{16pt}{\sffamily \bfseries \color{green!50!black}}{:}{.5em}{}
\theoremstyle{theor}

\usepackage[explicit]{titlesec}

\newcommand*\sectionlabel{}
\titleformat{\section}
  {\gdef\sectionlabel{}
   \Large\bfseries\scshape}
  {\gdef\sectionlabel{\thesection }}{0pt}
  {\begin{tikzpicture}[remember picture,overlay]
       \end{tikzpicture}
  }
\titlespacing*{\section}{0pt}{0pt}{0pt}

\newcommand*\subsectionlabel{}
\titleformat{\subsection}
  {\gdef\subsectionlabel{}
   \large\bfseries\scshape}
  {\gdef\subsectionlabel{\thesubsection  }}{0pt}
  {\begin{tikzpicture}[remember picture,overlay]
    	\draw (-0.15, 0.02) node[right] {\color{\stylecolor} \textsf{\subsectionlabel}};
	\draw (1.25, 0) node[right] {\color{\stylecolor} \textsf{#1}};
	\fill[color=\stylecolor] (1,-0.25) rectangle (1.1, 0.25);
       \end{tikzpicture}
  }
\titlespacing*{\subsection}{0pt}{10pt}{10pt}

\newcommand*\subsubsectionlabel{}
\titleformat{\subsubsection}
  {\gdef\subsubsectionlabel{}
   \bfseries\scshape}
  {\gdef\subsubsectionlabel{\thesubsubsection.\ \  }}{0pt}
  {\begin{tikzpicture}[remember picture,overlay]
    	\draw (-0.15, 0) node[right] {\color{\stylecolor} \textsf{\subsubsectionlabel#1}};
       \end{tikzpicture}
  }
\titlespacing*{\subsubsection}{0pt}{7pt}{7pt}

\pgfplotsset{every axis legend/.append style={at={(1.02,1)},anchor=north west}}

\begin{document}

\allowdisplaybreaks

\pagestyle{fancy}
\renewcommand{\headrulewidth}{0pt}
\fancyhead{}

\fancyfoot{}
\fancyfoot[C] {\textsf{\textbf{\thepage}}}

\begin{equation*}
\begin{tikzpicture}
\draw (\textwidth, 0) node[text width = \textwidth, right] {\color{white} easter egg};
\end{tikzpicture}
\end{equation*}

\begin{equation*}
\begin{tikzpicture}
\draw (0.5\textwidth, -3) node[text width = \textwidth] {\huge  \textsf{\textbf{Stokes paradox in electronic Fermi liquids}} };
\end{tikzpicture}
\end{equation*}
\begin{equation*}
\begin{tikzpicture}
\draw (0.5\textwidth, 0.1) node[text width=\textwidth] {\large \color{black} \textsf{Andrew Lucas}};
\draw (0.5\textwidth, -0.5) node[text width=\textwidth] {\small \textsf{Department of Physics, Stanford University, Stanford, CA 94305, USA}};
\end{tikzpicture}
\end{equation*}
\begin{equation*}
\begin{tikzpicture}
\draw (0, -13.1) node[right, text width=0.5\paperwidth] {\texttt{ajlucas@stanford.edu}};
\draw (\textwidth, -13.1) node[left] {\textsf{\today}};
\end{tikzpicture}
\end{equation*}
\begin{equation*}
\begin{tikzpicture}
\draw[very thick, color=\stylecolor] (0.0\textwidth, -5.75) -- (0.99\textwidth, -5.75);
\draw (0.12\textwidth, -6.25) node[left] {\color{\stylecolor}  \textsf{\textbf{Abstract:}}};
\draw (0.53\textwidth, -6) node[below, text width=0.8\textwidth, text justified] {\small The Stokes paradox is the statement that in a viscous two dimensional fluid, the ``linear response" problem of fluid flow around an obstacle is ill-posed.   We present a simple consequence of this paradox in the hydrodynamic regime of a Fermi liquid of electrons in two-dimensional metals.     Using hydrodynamics and kinetic theory, we estimate the contribution of a single cylindrical obstacle to the global electrical resistance of a material, within linear response.  Momentum relaxation, present in any realistic electron liquid, resolves the classical paradox.   Nonetheless, this paradox imprints itself in the resistance, which can be parametrically larger than predicted by Ohmic transport theory.    We find a remarkably rich set of behaviors, depending on whether or not the quasiparticle dynamics in the Fermi liquid should be treated as diffusive, hydrodynamic or ballistic on the length scale of the obstacle.  We argue that all three types of behavior are observable in present day experiments.};
\end{tikzpicture}
\end{equation*}

\tableofcontents

\titleformat{\section}
  {\gdef\sectionlabel{}
   \Large\bfseries\scshape}
  {\gdef\sectionlabel{\thesection }}{0pt}
  {\begin{tikzpicture}[remember picture,overlay]
	\draw (1, 0) node[right] {\color{\stylecolor} \textsf{#1}};
	\fill[color=\stylecolor] (0,-0.35) rectangle (0.7, 0.35);
	\draw (0.35, 0) node {\color{white} \textsf{\sectionlabel}};
       \end{tikzpicture}
  }
\titlespacing*{\section}{0pt}{15pt}{15pt}

\begin{equation*}
\begin{tikzpicture}
\draw[very thick, color=\stylecolor] (0.0\textwidth, -5.75) -- (0.99\textwidth, -5.75);
\end{tikzpicture}
\end{equation*}

\section{Introduction and Summary of Results}
One of the simplest experimental probes of a solid state system is to calculate the electrical resistance of a sample.   Despite the experimental simplicity, theories of transport often rely on simplistic and unjustified assumptions.    For example, common lore states that electrical resistance $\rho \sim T^2$ in an ordinary metal, since quasiparticles near the  Fermi surface can scatter at a rate $T^2$.   In fact, $\rho$ is sensitive not to the mean free path of quasiparticles, but to the rate of momentum relaxation \cite{lucasMM}.   

Experimental data from non-Fermi liquids, including the cuprates \cite{vandermarel, hussey} and charge-neutral graphene \cite{crossno, lucas3}, demand a more rigorous theory of transport.    But due to the challenge of quantitatively computing $\rho$  in these strongly interacting systems, it is important to look for new organizing principles for a theory of transport.  One exciting possibility is that the flow of electrons is hydrodynamic \cite{lucas3, hkms, andreev}.  On general grounds, we expect an interacting classical or quantum system to thermalize, and that the dynamics of this thermalization on long time and length scales will be governed by hydrodynamics \cite{kadanoff, landau}.   In practice, this hydrodynamic regime is hard to see directly in experiment: electrons usually scatter frequently off of impurities and/or phonons, which can reduce their dynamics to single-particle dynamics in a disordered background.

In this paper, we will focus on the consequences of hydrodynamic electron flow in Fermi liquids.   These weakly interacting quantum fluids consist of long-lived quasiparticles:  as such, Fermi liquid theory is under quantitative control \cite{pines}.   Although Fermi liquids are weakly interacting, they are still interacting:  in a very clean crystal, the electronic Fermi liquid will exhibit hydrodynamic behavior.   Unlike the non-Fermi liquids mentioned previously, however, the viscosity and thermodynamics of a Fermi liquid remain under better theoretical control, making them ideal settings for comparing hydrodynamic theory to experimental data.   Evidence for hydrodynamic electron flow has been observed in Fermi liquids in multiple different materials \cite{molenkamp, tauk, bandurin, mackenzie}.   Recent theoretical work has demonstrated a variety of interesting hydrodynamic phenomena in Fermi liquids, including novel local \cite{andreev, alekseev} and nonlocal \cite{polini, levitovhydro, levitov1607} transport signatures.  Signatures of nonlinear hydrodynamics may also be possible \cite{succiturb, tomadin}.   

Of particular interest to us will be to find an experiment which can distinguish between three distinct regimes of electronic dynamics:  \begin{itemize}
\item \textbf{diffusive:}  Quasiparticles rapidly scatter off of impurities and/or phonons, and quasiparticle-quasiparticle scattering is negligible.  Most dynamical degrees of freedom can be integrated out:  electrical transport is well described by the diffusion of the conserved electronic charge.
\item \textbf{ballistic:}  Quasiparticle-quasiparticle scattering remains negligible, but quasiparticle-impurity/phonon scattering is comparable to other length scales in the problem, such as the material length.   Few dynamical degrees of freedom can be integrated out:  a fully kinetic description is required.
\item \textbf{hydrodynamic:} Quasiparticle-quasiparticle scattering is fast compared to quasiparticle-impurity/phonon scattering, and compared to the travel time across the sample.  Most degrees of freedom can be integrated out;  however, we must account for the motion of conserved charge and approximately conserved momentum (and possibly energy). 
\end{itemize}
Although much effort has gone into distinguishing the diffusive and hydrodynamic regimes, it is also important to distinguish the hydrodynamic and ballistic regimes.   As Fermi liquids where hydrodynamics has been observed to date are either two dimensional, or effectively two dimensional, we will focus on this dimension henceforth.

\subsection{Stokes Paradox}
 One particularly striking phenomenon in classical two dimensional fluid mechanics is the Stokes paradox:  if we try to push a fluid around a circular obstacle of radius $R$ with a small velocity $v_0$, the force the obstacle exerts on the fluid is not proportional to $v_0$ in the linear response limit $v_0 \rightarrow 0$ \cite{lamb}:\begin{equation}
 F \approx \dfrac{4\mpi \eta v_0}{\displaystyle \log \frac{\nu }{v_0 R}},
 \end{equation}
 with $\nu$ the dynamical viscosity and $\eta$ the shear viscosity.
This breakdown of linear response theory is a natural `paradox' of interest for two-dimensional electronic fluids:  so far all of the observations of electronic hydrodynamics have been in  linear response measurements.     As we will show in this paper, momentum-relaxing processes resolve this paradox in electronic fluids: linear response measurements such as the net electrical resistance $\mathcal{R}$ will be strictly finite.   However, the fact that electron-electron collisions do not relax momentum can have dramatic consequences for $\mathcal{R}$.   

\begin{figure}
\centering
\includegraphics[width=3.5in]{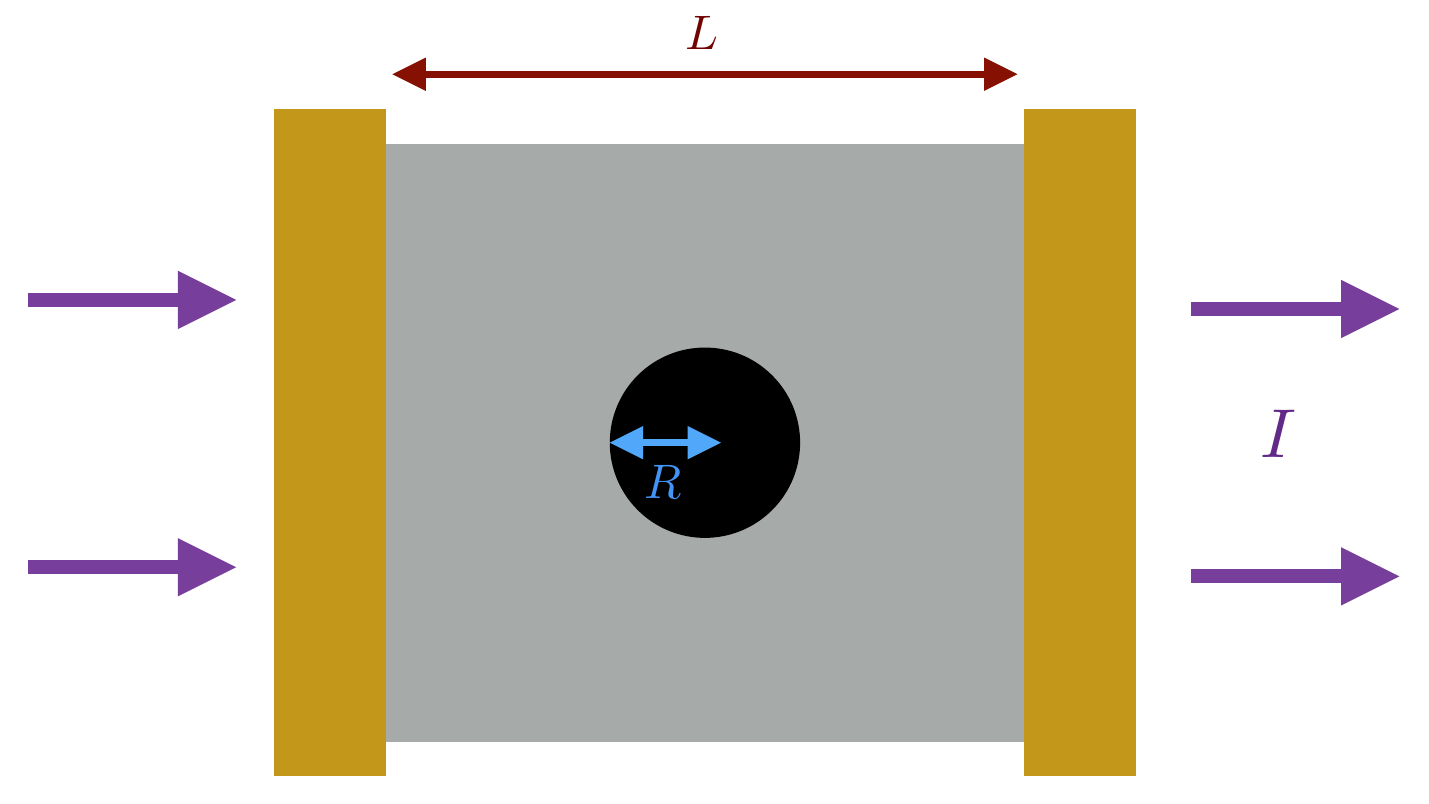}
\caption{A two-point electrical measurement for the resistance of a two-dimensional Fermi liquid, in a sample of size $L$.   A `hard' obstacle of radius $R\ll L$ is placed in the middle of the sample.}
\label{fig:setup}
\end{figure}

In this paper, we imagine studying an electronic analogue Stokes paradox in a two dimensional Fermi liquid via the simple set-up shown in Figure \ref{fig:setup}.   The global electrical resistance $\mathcal{R}$ is computed through a simple two-point electrical measurement.    Through gating or other mechanisms, a circular obstacle of radius $R$ is placed in the center of the sample;  we assume $R\ll L$, with $L$ the length/width of the sample.   This obstacle will enhance the electrical resistance $\mathcal{R}$.

Let us begin by imagining that the metal is an ordinary Ohmic, diffusive metal.  We will shortly discuss when the metal will behave this way.   But if it does, then we can simply compute the resistance of the metal by solving the Laplace equation with suitable boundary conditions.  If $R\ll L$,  then the resistance of the metal with the obstacle imposed is \begin{equation}
\mathcal{R}_{\mathrm{tot}} = \mathcal{R}_0 + \mathcal{R}^\prime
\end{equation}
with $\mathcal{R}_0$ the resistance of the obstacle-free metal, and
 \begin{equation}
\mathcal{R}^\prime  \approx c \mathcal{R}_0 \frac{R^2}{L^2}. \label{eq:expR}
\end{equation}
$c>0$ is an O(1) constant depending on the precise geometry and boundary conditions used in the experimental setup -- it could be found by solving Laplace's equation numerically.     Hence, we can think of the obstacle as approximately adding a ``series resistor" of resistance $\mathcal{R}^\prime$.  

As we will see, in order for (\ref{eq:expR}) to hold, we need $R$ to satisfy certain constraints.   First, we require $k_{\mathrm{F}}R\gg 1$:  this is the simple statement that the quasiparticle dynamics can be treated via semiclassical equations like the Ohmic diffusion equation.   Of interest in this paper are semiclassical length scales:  the mean free path between electron-electron collisions $\ell_0$, and between electron-impurity/phonon collisions $\xi$.  We will see that (\ref{eq:expR}) holds only when $R\gtrsim \sqrt{\ell_0\xi}$.

In the absence of momentum-relaxing collisions, it is a theorem that the electrical resistivity $\rho = 0$.   Indeed, if momentum is conserved, then we may boost to a moving reference frame where the global momentum of the Fermi liquid, and hence the electrical current, are finite.   We may do so without an applied electric field, and so this finite electrical current is consistent with Ohm's Law $E=\rho J$ only if $\rho=0$.   When (weak) momentum relaxing collisions are accounted for, the electrical resistivity is non-zero, and of Drude form:  \begin{equation}
\rho = \frac{m_0 }{n_0e^2}  \frac{v_{\mathrm{F}}}{\xi} \label{eq:Druderho}
\end{equation}
Here $e$ is the quasiparticle charge, $v_{\mathrm{F}}$ is the Fermi velocity, $m_0$ is the quasiparticle mass and $n_0$ is the quasiparticle density.    It is crucial to note that $\rho$ depends on $\xi$, but not on $\ell_0$.  Since $\mathcal{R}_0\sim \rho$, transport measurements in the absence of the obstacle give us a clean experimental measurement of $\xi$ (if quasiparticles are well-defined).  

Our main result is that the electrical resistance of the metal (with the obstacle) depends not only on $\xi$, but on $R$ and $\ell$ as well.   This dependence is rather complicated.  We will present a heuristic argument that $\mathcal{R}^\prime$ becomes enhanced to $\mathcal{R}^\prime  = c\mathcal{R}_0 R_{\mathrm{eff}}^2/L^2$:  the obstacle has an effective radius $R_{\mathrm{eff}}$ given by \begin{equation}
R_{\mathrm{eff}}^2 \approx \ell \xi \left[\left(1-\frac{2\ell}{\xi}\right)\log \left(\frac{\xi}{2\ell}\left(\sqrt{1+\left(\frac{2\ell}{R}\right)^2}-1\right) + 1\right) + \sqrt{1+\left(\frac{2\ell}{R}\right)^2}-1\right]^{-1},
 \label{eq:main}
\end{equation}
where we have defined \begin{equation}
\frac{1}{\ell} \equiv \frac{1}{\ell_0} + \frac{1}{\xi}.  \label{eq:mattheisen}
\end{equation}
Using simple hydrodynamic and kinetic models of two dimensional Fermi liquids with circular Fermi surfaces, we will confirm numerically that this result is surprisingly accurate, predicting the true value of $R_{\mathrm{eff}}$ to $\lesssim$30\% (at least, for every set of parameters we can test, which includes all feasible experimental parameters).  We expect that qualitative features of (\ref{eq:main}) are robust against the Fermi surface shape.  For the remainder of the introduction, we will take (\ref{eq:main}) as an `exact' result, and explore its phenomenological consequences.    

There are three limits of interest of (\ref{eq:main}):  \begin{itemize}
\item \textbf{diffusive:}  in the limit when $R\gg \sqrt{\ell\xi}$, we find \begin{equation}
R_{\mathrm{eff}}^2 \approx R^2.  \label{eq:maindif}
\end{equation}

\item \textbf{ballistic:} in the limit when $\ell \gg R$, we find \begin{equation}
R_{\mathrm{eff}}^2 \approx \frac{\xi R}{2}  \label{eq:mainbal}
\end{equation}
if $\xi \sim \ell$, and \begin{equation}
R_{\mathrm{eff}}^2 \approx \dfrac{\xi}{\displaystyle \frac{2}{R} + \frac{1}{\ell} \log \frac{\xi}{R}}. \label{eq:mainbal2}
\end{equation} if $\xi \gg \ell$.   While in principle, if $\xi \rightarrow \infty$ the logarithmic term in the denominator can dominate, this would require $\xi$ (and hence the conductivity of the obstacle-free sample) to be exponentially large,  which is likely impossible in practice.   (\ref{eq:mainbal}) admits a simple explanation.  If $\xi \sim \ell >R$, then there is a forbidden region of length $\sim \xi$ and width $R$ in front/behind the obstacle where particles ballistically scatter and propagate against the direction of the current.  Hence, the effective area blocked by the obstacle is $R_{\mathrm{eff}}^2 \sim \xi \times R$.  This effect can be observed numerically in Figure \ref{fig:scatter}.

\item \textbf{hydrodynamic:}  in the limit when $\ell \ll R \ll \sqrt{\ell\xi}$, we find \begin{equation}
R_{\mathrm{eff}}^2 \approx \dfrac{\ell \xi}{\displaystyle \log\frac{ \ell\xi}{R^2} }.   \label{eq:mainhydro}
\end{equation}
Interestingly, $R_{\mathrm{eff}}$ is only logarithmically independent on the actual radius $R$.   To understand (\ref{eq:mainhydro}), imagine a `colored' random walker which can step a distance $\sim \ell$, conserving its `color' (momentum) at almost every collision.   This color is sometimes altered during a collision, with probability $p=\ell/\xi \ll 1$, \emph{or} if the random walker hits the region $r\le R$.   Now we ask: starting at $r=R$ at $t=0$,  how far from the origin does a colored walker typically get before losing its color?   Neglecting the collisions at $r=R$,  we would expect that $\ell^2/p \sim \ell \xi \sim r^2$.   But in two dimensions, random walks repeatedly return to their starting point \cite{redner}:  at time $t$ the probability of being found at $r=R$ is $\sim t^{-1}$.   Integrating over $t$, the number of collisions at $r=R$ scales as $\log t(r) \sim \log (r/R)$.   So we modify our previous estimate, crudely rescaling $p\rightarrow p/\log(r/R)$:   hence we estimate $\ell^2/p \times \log(r/R) \sim r^2$.   To leading logarithmic order, the solution is $r=R_{\mathrm{eff}}$, with $R_{\mathrm{eff}}$ given by (\ref{eq:mainhydro}).   Since the walker must travel this far before ``losing" its momentum,  we identify this as the effective size of the obstacle.
\end{itemize}
(\ref{eq:maindif}) and (\ref{eq:mainhydro}) can be shown rigorously.  Note that in the hydrodynamic and ballistic limits, $R_{\mathrm{eff}} \gg R$.  It is not difficult to prove that $R_{\mathrm{eff}}\ge R$ for any values of $\xi$ and $\ell$, and so the obstacle always appears \emph{bigger} than it truly is.    In other words,  $\mathcal{R}^\prime$ is always larger than it is if transport can be described by Ohmic, diffusive transport.

\subsection{Predictions for Fermi Liquids}
The model with which we derive (\ref{eq:main}) is a model of two-dimensional Fermi liquids.  So let us now analyze (\ref{eq:main}), employing Fermi liquid phenomenology, and determine the temperature dependence of $\mathcal{R}^\prime$.    In a Fermi liquid, we expect that in the absence of impurities \cite{pines} \begin{equation}
\ell_0 = \mathcal{C} \frac{\hbar e \mu_* v_{\mathrm{F}}}{(k_{\mathrm{B}}T)^2} \equiv R \left(\frac{T_*}{T}\right)^2,  \label{eq:FLell}
\end{equation}
where $\mu$ is the chemical potential, $v_{\mathrm{F}}$ is the Fermi velocity, and $\mathcal{C}$ is a dimensionless constant (likely of order unity, unless the Fermi surface is rather oddly shaped).  This equation will hold so long as $e \mu_* \gg k_{\mathrm{B}}T$.    $T_*$ is the temperature at which $\ell = R$ in the absence of impurity scattering:  $k_{\mathrm{B}}T_* = \sqrt{\mathcal{C}\hbar \mu_*v_{\mathrm{F}}/R}$.    Logarithmic corrections to (\ref{eq:FLell}) are expected for a two dimensional Fermi liquid \cite{logT2, polini1506}.   These corrections multiply $\mathcal{C}$ by a power of $\log[\mu/T]$, which may be a numerically non-negligible prefactor, but will not modify the fact that $\ell$ is a rapidly decreasing function of $T$ in a Fermi liquid.   For the qualitative discussion which follows below, the precise functional form of $\ell$ is not crucial.

\begin{figure}
\centering
\includegraphics{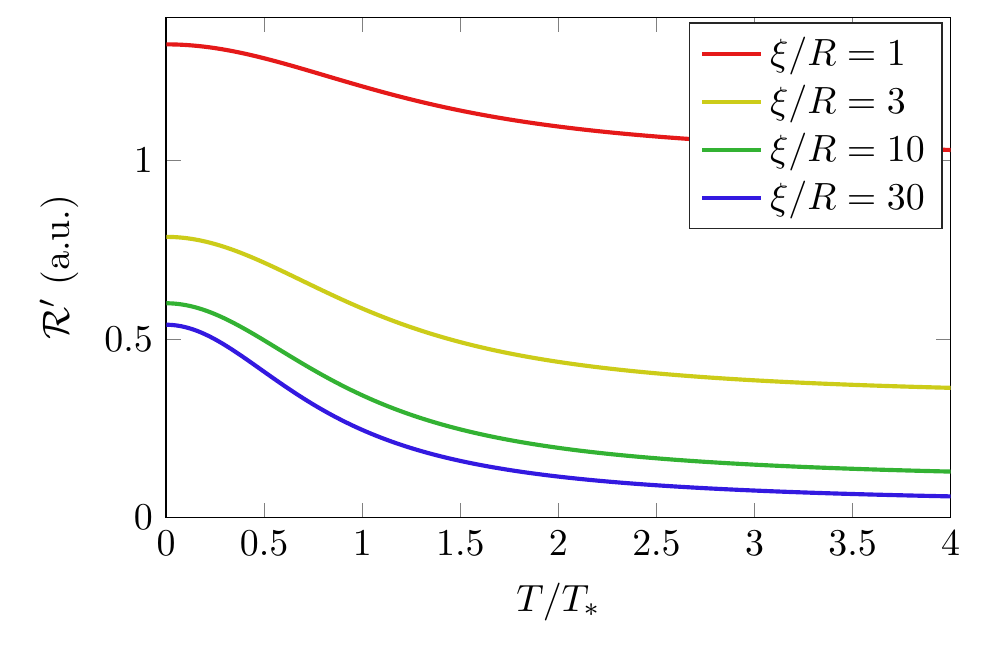}
\caption{The temperature dependence of the obstacle resistance $\mathcal{R}^\prime$, neglecting electron-phonon scattering.  The overall resistance $\mathcal{R}^\prime$ decreases as $\xi$ increases (as expected from (\ref{eq:Druderho})).  Once $\xi \gtrsim R$,  $\mathcal{R}^\prime$ exihibits a sharp drop as temperature increases.  This can equivalently be characterized by a large spike in $R_{\mathrm{eff}}$ at low $T$.}
\label{fig:FL1}
\end{figure}

Let us begin by supposing that $\xi$ comes from electron-impurity scattering and is approximately temperature independent.  Figure \ref{fig:FL1} shows the temperature dependence of $\mathcal{R}^\prime \sim R_{\mathrm{eff}}^2/\xi$ in this limit.   We can see that so long as $\xi \gtrsim 10R$, there is a nearly order-of-magnitude difference between the low temperature and high temperature limit.   Furthermore, the drop in $\mathcal{R}^\prime$ occurs fairly rapidly.   The dramatic temperature dependence of Figure \ref{fig:FL1} is a direct consequence of the hydrodynamic limit.   
 
\begin{figure}
\centering
\includegraphics{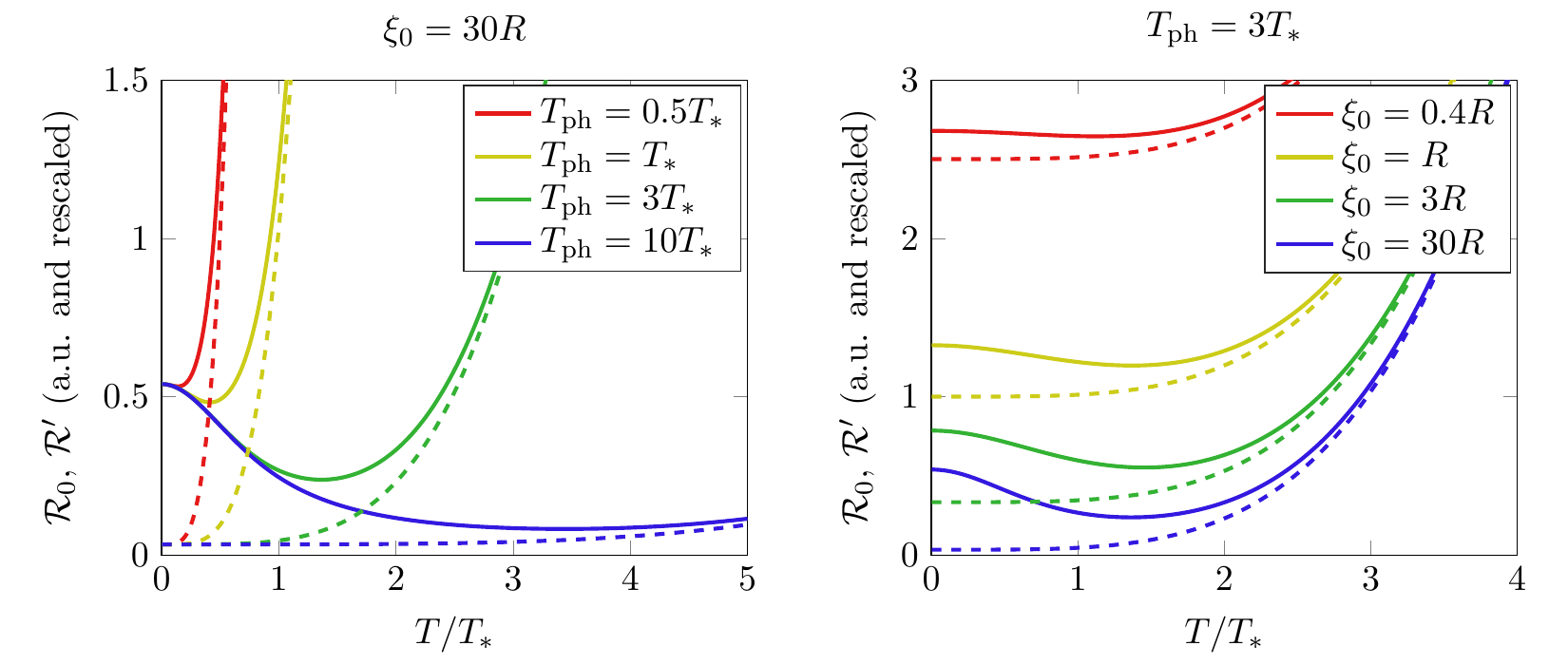}
\caption{The temperature dependence of the obstacle resistance $\mathcal{R}^\prime$, and the sample resistance $\mathcal{R}_0$, accounting for temperature dependent electron-phonon scattering.    Solid lines plot $\mathcal{R}^\prime$, and dashed lines plot $\mathcal{R}_0$.   We have rescaled $\mathcal{R}_0$ and $\mathcal{R}^\prime$ so that the solid and dashed curves will coincide in the diffusive limit $R_{\mathrm{eff}}\approx R$.    In the left panel, we see the consequences of the ratio $T_{\mathrm{ph}}/T_*$ at fixed $\xi_0$.   Unless $T_{\mathrm{ph}}>T_*$, it is impossible to see any clear evidence of hydrodynamic flow.    In the right panel, we see that a weak signature for hydrodynamic flow persists down to surprisingly small $\xi_0 \sim R$, once $T_{\mathrm{ph}}\gtrsim T_*$. }
\label{fig:FL2}
\end{figure}

The rapid nature of this drop is important.  In a real metal, electron-phonon scattering will inevitably occur at higher temperatures, and so $\xi$ will also be temperature dependent.   In the low temperature limit, on general principles we expect that the (acoustic) phonon contribution to the resistivity scales as $T^4$ \cite{hwang07, efetov},  and so we make the ansatz \begin{equation}
\frac{1}{\xi} = \frac{1}{\xi_0} + b T^4 =  \frac{1}{\xi_0} + \frac{1}{R} \left(\frac{T}{T_{\mathrm{ph}}}\right)^4,
\end{equation}
where $\xi_0$ is the electron-impurity scattering length, which we assume is temperature-independent,  $b$ is a microscopic parameter which is $R$-independent, and $T_{\mathrm{ph}} \equiv (bR)^{-1/4}$ is the temperature at which electron-phonon scattering is fast enough to render $\xi = R$ in an impurity-free sample.   $\ell$ is still given by (\ref{eq:mattheisen}) and (\ref{eq:FLell}), and so $\ell$ depends on both $T_*$ and $T_{\mathrm{ph}}$.    To determine whether or not this spoils the clear signature of hydrodynamic electron flow seen in Figure \ref{fig:FL2}, we compare both $\mathcal{R}_0$ and $\mathcal{R}^\prime$ in Figure \ref{fig:FL2}, for a variety of different ratios $\xi_0/R$ and $T_{\mathrm{ph}}/T_*$.    As is clear from the figure, the crucial signature of hydrodynamics is not just the $T$-dependence of $\mathcal{R}^\prime$:  it is the \emph{decrease} of $\mathcal{R}^\prime$ over intermediate temperatures.   To see this signal, the ratio $T_{\mathrm{ph}}/T_* \gtrsim 2$ seems necessary.  Indeed, if $T_{\mathrm{ph}} < T_*$ then it is impossible to get into the regime $\ell \ll \xi$, which is necessary to observe hydrodynamic physics.   

Simply seeing a large enhancement of $R_{\mathrm{eff}}$ is \emph{not} sufficient evidence for hydrodynamic flow.  Indeed, from (\ref{eq:mainbal2}),  $R_{\mathrm{eff}}$ is actually largest in the ballistic regime.       In particular, if we take $\ell = \xi$,  then the differential resistance \begin{equation}
\mathcal{R}^\prime \sim \frac{R_{\mathrm{eff}}^2}{\xi} \sim \frac{\xi}{\displaystyle \sqrt{1+\left(\dfrac{2\xi}{R}\right)^2} -1 - \log\left(\dfrac{1}{2}+\dfrac{1}{2}\sqrt{1+\left(\dfrac{2\xi}{R}\right)^2}\right)}.
\end{equation}
One can straightforwardly check that $\mathcal{R}^\prime$ is a decreasing function of $\xi$.   Hence, if there is no separation of length scales between $\xi$ and $\ell$,  then so long as one observes that $\partial \rho / \partial T > 0$  (and hence $\partial \xi/\partial T<0$),  one must also observe $\partial \mathcal{R}^\prime/\partial T>0$.   We re-iterate that it is the exotic non-monotonic temperature dependence of $\mathcal{R}^\prime$:   $\partial \mathcal{R}^\prime/\partial T <0$ even as $\partial \rho /\partial T > 0$, which signals the hydrodynamic limit.

\subsection{Comparison with Older Results}
Let us briefly compare our hydrodynamic results with older results \cite{levitov1607, spivak02, succicond}, which also discuss hydrodynamic contributions to resistance due to scattering off of point-like impurities.   In this earlier literature, the electron fluid is modeled as perfectly translation invariant, up to small point-like impurities; hence $\xi = \infty$.   The more microscopic computation of \cite{levitov1607} gives \begin{equation}
\mathcal{R}^\prime \sim \dfrac{\ell}{\displaystyle \log \frac{L_0}{\min(R, \sqrt{\ell R})}}   \label{eq:old2}
\end{equation}
when $L_0 \rightarrow \infty$.   \cite{spivak02} finds a similar result, but with incorrect $\ell$-dependence of the logarithm.   In a Fermi liquid, this viscous-dominated theory of transport leads to a dramatic experimental prediction:  $\rho \sim \ell \sim T^{-2}$ (up to logarithmic corrections) in the hydrodynamic regime $\ell \ll \xi$.   This effect has not been experimentally observed to date in a Fermi liquid, even in Fermi liquids  where other signatures for hydrodynamics have been observed.   As such, we expect that the dominant contribution to the electrical resistivity of a typical sample (without such obstacles added by hand) is coming from other kinds of electron-impurity/phonon and/or umklapp scattering.   In our model, such effects are accounted for by the parameter $\xi$, and as we will see, this leads to important qualitative changes.

First, from (\ref{eq:main}) and (\ref{eq:Druderho}), we note that $\mathcal{R}^\prime = \mathcal{A} R_{\mathrm{eff}}^2 / \xi$, where $\mathcal{A}$ is a geometric prefactor which does not depend on any of the semiclassical length scales $R$, $\ell$, $\xi$.    Using (\ref{eq:mainhydro}), we obtain (when $ R \ll \sqrt{\ell\xi}$) \begin{equation}
\mathcal{R}^\prime = \mathcal{A} \dfrac{\ell}{\displaystyle \log\left(\frac{\xi}{R}\min\left(1,\frac{\ell}{R}\right)\right) }.  \label{eq:new}
\end{equation}
Both (\ref{eq:old2}) and (\ref{eq:new})  have the same $\ell$-dependence outside of the logarithm (and hence the same viscosity and temperature dependence).   However, in (\ref{eq:new}) the UV scale at which the Stokes paradox ends is set `by hand' to $L_0$ in the computations of \cite{spivak02, levitov1607}, whereas it is $\sqrt{\ell\xi}$ in our model, even when there is only a single cylindrical impurity.    Unlike the models of \cite{spivak02, levitov1607}, the Stokes paradox has been `cured' by $\xi < \infty$.   The argument of the logarithm in (\ref{eq:new}) agrees with (\ref{eq:old2}) upon setting $L_0 \sim \sqrt{\ell \xi}$, the length scale at which the Ohmic description of transport is valid in our model.   

As we send $\ell\rightarrow 0$ (at fixed $R$ and $\xi$), once $\sqrt{\ell \xi} \lesssim R$, the viscous mode no longer contributes to transport.   Instead, $\mathcal{R}^\prime$ is given by the diffusive answer:  $\mathcal{R}^\prime = \mathcal{A}R^2/\xi$.   In the regime $\sqrt{\ell \xi} \gg R$ where viscous effects matter,   $ \mathcal{A} R^2/\xi   \ll \mathcal{A}\ell / \log(\sqrt{\ell \xi}/R)$:   the viscous mode thus \emph{enhances} the conductivity over what it would be if transport was described by Ohmic diffusion.     From (\ref{eq:mainbal2}),  we see that if $\xi \gg \ell \gg R$, viscosity \emph{reduces} $\mathcal{R}^\prime$ over what it would be if obstacle scattering was ballistic.   The computations of \cite{levitov1607, spivak02} only note this viscous reduction of transport over the ballistic regime.   In fact, the Stokes paradox also serves to enhance $\mathcal{R}^\prime$ above a diffusion-limited value:  indeed, when $\xi = \infty$, the diffusion-limited value of $\mathcal{R}^\prime = 0$, and so this effect is invisible.

\subsection{Outline}
The remainder of this paper is as follows.    In Section \ref{sec2}, we discuss the resolution of the Stokes paradox within momentum-relaxing hydrodynamics.  In Section \ref{sec3}, we generalize to a kinetic theory capable of describing the ballistic-to-hydrodynamic crossover, borrowing techniques from the elegant paper \cite{levitov1607}.   We conclude in Section \ref{sec4} by discussing methods of implementing our setup in present day experiments.   

Sections \ref{sec2} and \ref{sec3} are written in a lengthy and pedagogical way, with intermediate steps of calculations often shown, and connections between kinetic theory and hydrodynamics emphasized.   These sections may be skipped by readers who are not interested in technicalities.

\section{Hydrodynamics}\label{sec2}
We begin with a hydrodynamic treatment of the Stokes paradox in an electronic Fermi liquid of chemical potential $\mu$, with  $e\mu \gg k_{\mathrm{B}}T$.    Hydrodynamics is an effective description valid on length scales large compared to $\ell$ \cite{kadanoff, landau}; the equations of motion are conservation laws for locally conserved quantities: charge, energy and momentum.   For simplicity, we neglect energy conservation, as we expect thermal effects to negligibly correct the solutions below when $\mu \gg T$.   Charge conservation reads \begin{equation}
\partial_i (n v_i ) = 0  \label{eq:hydro1}
\end{equation}
with $n$ the local charge density.  The momentum conservation equation (Navier-Stokes equation) must be modified to account for the (weak) effects of momentum relaxation out of the electronic fluid due to impurities.   We do so in a simple ``mean field" way: \begin{equation}
n v_j \partial_j v_i + \partial_i P - \partial_j \left( \eta \left(\partial_i v_j + \partial_j v_i - \mdelta_{ij} \partial_k v_k \right)\right) = -en\partial_i \varphi -\Gamma v_i  \label{eq:hydro2}
\end{equation}
$\Gamma$ characterizes the rate of momentum loss due to impurities, and can be computed in a specific microscopic model following the methods of \cite{lucasMM}.  $P$ is the pressure and $\varphi$ is the electric potential due to long range Coulomb interactions.   $\varphi$ enters the equations of motion due to a Lorentz force acting on the charge carriers.

If we neglect thermal fluctuations, we may use the thermodynamic identity \begin{equation}
\mathrm{d}P = en\mathrm{d}\mu  \label{eq:dP},
\end{equation}
to re-write (\ref{eq:hydro2}) in terms of the variable $\tilde \mu = \mu + \varphi$: \begin{equation}
n v_j \partial_j v_i + en \partial_i \tilde \mu - \partial_j \left( \eta \left(\partial_i v_j + \partial_j v_i - \mdelta_{ij} \partial_k v_k \right)\right) =  -\Gamma v_i
\end{equation}
 In terms of the variable $\tilde \mu$,  (\ref{eq:hydro2}) is \emph{identical} to the hydrodynamic equations governing transport in a fluid without long range Coulomb interactions ((\ref{eq:hydro2}) with $\varphi=0$).    As emphasized in \cite{lucasMM, lucas3},   our direct current transport experiment of the conductivity measures $\tilde \mu$.     Henceforth, we drop the tilde on $\mu$, absorbing the effects of long range Coulomb interactions.

\subsection{Linear Response}
Let us now consider the linear response limit of (\ref{eq:hydro1}) and (\ref{eq:hydro2}).  We expand around a static fluid with $v_i=0$, and background number density $n\approx n_0$.   To linear order, (\ref{eq:hydro1}) becomes \begin{equation}
\partial_i v_i = 0,  \label{eq:incomp}
\end{equation}
which means that we may write the velocity in terms of a stream function: \begin{equation}
v_i = \epsilon_{ij}\partial_j \psi,
\end{equation}
with $\epsilon_{ij}$ the Levi-Civita tensor.   By applying $\epsilon_{ij}\partial_j$ to (\ref{eq:hydro2}), we find \begin{equation}
\partial_i \partial_i \partial_j \partial_j \psi = \frac{1}{\lambda^2}\partial_i \partial_i \psi.  \label{eq:4ord}
\end{equation}
where we have defined the hydrodynamic momentum relaxation length \begin{equation}
\lambda \equiv \sqrt{\frac{\eta}{\Gamma}}.  \label{eq:lambda}
\end{equation}

A sufficient family of solutions to (\ref{eq:4ord}) for our purposes is \begin{equation}
\psi = \psi_1+\psi_2,   \label{eq:psi12}
\end{equation}
where\footnote{If this were a finite-dimensional linear problem, this class of solutions would be complete.  This follows from the linear algebra result that if $\mathsf{AB}\mathbf{v}=\mathbf{0}$ and $\mathsf{AB}=\mathsf{BA}$, then $\mathbf{v}\in \mathrm{ker}(\mathsf{A})\oplus \mathrm{ker}(\mathsf{B})$.   This statement is easily proved by finding a basis in which $\mathsf{A}$ and $\mathsf{B}$ are simultaneously diagonal.} \begin{subequations}\begin{align}
\partial_i \partial_i \psi_1 &= 0, \\ 
\partial_i \partial_i \psi_2 &= \frac{\psi_2}{\lambda^2}.
\end{align}\end{subequations}
We recognize that $\psi_2$ is a ``gapped" or ``massive" degree of freedom -- regularity at infinity means that $\psi_2$ will only be non-negligible over the scale $\lambda$.    For physics on long length scales compared to $\lambda$, we will find (to exponential accuracy)  $\psi \approx \psi_1$.   

This simple observation has very important consequences for experiment.   If $\psi_2$ is negligible, then this is equivalent to the approximation that \begin{equation}
\partial_i \mu \approx -\Gamma v_i.  \label{eq:preohm}
\end{equation}
Combining  (\ref{eq:dP}),  (\ref{eq:hydro1}) and (\ref{eq:preohm}) we conclude that \begin{equation}
\partial_i \left(-\frac{1}{\rho} \partial_i \mu\right) = 0,  \label{eq:ohm}
\end{equation}
where \begin{equation}
\rho \equiv \frac{\Gamma}{e^2 n_0^2}  \label{eq:rho}
\end{equation}
 is precisely the Drude resistivity \cite{lucasMM}.   On length scales long compared to $\lambda$, this hydrodynamic system reduces to a simple diffusive model, and it seems as though there is no possible experiment that could detect signatures of hydrodynamics.   But as we have already pointed out in (\ref{eq:main}),  if we have flow around a (circular) obstacle with a width small compared to $\lambda$,  the effective size of the disk can be modified substantially due to the change in boundary conditions at the obstacle boundary.    We will shortly confirm (\ref{eq:main}) explicitly.  

Let us make two final observations of interest.   Writing (\ref{eq:hydro2}) as \begin{equation}
en_0\partial_i \mu = \Gamma \epsilon_{ij} \partial_j \left(\lambda^2 \partial_k \partial_k \psi - \psi\right),
\end{equation} we see that \begin{equation}
en_0\partial_i \mu = - \Gamma \epsilon_{ij}\partial_j \psi_1.  \label{eq:mupsi1}
\end{equation}
Namely, the diffusive contribution $\psi_1$ to the stream function is the only one which can contribute at all to chemical potential (and hence voltage) shifts in the fluid, at least in our hydrodynamic model.

\subsection{No-Slip Boundary Conditions}
We now turn to the solution of the Stokes paradox in momentum relaxing hydrodynamics.    We wish to look for a solution which asymptotes to \begin{equation}
\mathbf{v} = u \hat{\mathbf{x}}
\end{equation}
subject to the boundary conditions that $v_i(r=R)=0$.    By rotational symmetry, we know that $\psi = f(r)\sin\theta$.   By writing $\psi$ as (\ref{eq:psi12}) is straightforward to show that \begin{equation}
\psi = \sin \theta \left[ Ar + \frac{B}{r} + C\mathrm{I}_1 \left(\frac{r}{\lambda}\right) + D\mathrm{K}_1 \left(\frac{r}{\lambda}\right)\right].  \label{eq:genhydro}
\end{equation}

To proceed further, we must impose boundary conditions.  We begin by imposing no-slip boundary conditions at $r=R$:  \begin{equation}
v_i(r=R) = 0.
\end{equation}
The boundary condition that $v_r = 0$ follows from the fact that charge cannot flow into the obstacle; the boundary condition $v_\theta=0$ implies `friction' between the obstacle and the fluid at the microscopic scale.
The boundary conditions at $r=\infty$ fix $A=u$ and $C=0$.   The boundary conditions at $r=R$ impose \begin{subequations}\begin{align}
0 &= uR + \frac{B}{R} + D\mathrm{K}_1\left(\frac{R}{\lambda}\right),  \label{eq:vr0} \\
0 &= u - \frac{B}{R^2} + \frac{D}{\lambda}\mathrm{K}_1^\prime\left(\frac{R}{\lambda}\right).
\end{align}\end{subequations}
After using Bessel function identities \cite{abramowitz}, one finds the final answer \begin{equation}
\psi = u\sin\theta \left[r - \left(R+2\lambda\frac{\mathrm{K}_1(R/\lambda)}{\mathrm{K}_0(R/\lambda)}\right)\frac{R}{r} + 2\lambda\frac{\mathrm{K}_1(r/\lambda)}{\mathrm{K}_0(R/\lambda)}\right].  \label{eq:noslipfinal}
\end{equation}

Let us now consider the leading corrections to this velocity as $r\rightarrow \infty$.   When $R\gg \lambda$, we find \begin{equation}
\psi \approx u\sin\theta\left[r-\frac{R^2}{r}\right],  \label{eq:asy1}
\end{equation}
except for $r\sim R$.   This is precisely the solution corresponding to Ohmic flow.   So without voltage probes very close to the surface of the obstacle, there is no way to detect hydrodynamic flow.   However, if $R\ll \lambda$, we instead find \begin{equation}
\psi(r\gg \lambda)\approx u\sin\theta\left[r-\frac{2\lambda^2}{\log \frac{\lambda}{R}} \frac{1}{r}\right].  \label{eq:asy2}
\end{equation}
By comparing to the Ohmic response (\ref{eq:asy1}), we conclude that from far away, the cylinder appears to be \emph{larger}: \begin{equation}
R_{\mathrm{eff}}^2 \approx \frac{2\lambda^2}{\log \frac{\lambda}{R}}.  \label{eq:Reff}
\end{equation} 

Following (\ref{eq:mupsi1}), we may now compute (for any ratio $R/\lambda$) \begin{equation}
\mu = - \rho J \cos\theta \left(r+\frac{B}{r}\right).   \label{eq:mudiff}
\end{equation}
We have defined \begin{equation}
J \equiv en_0u 
\end{equation}
to be the electric current density at $r=\infty$.    One can readily see that (\ref{eq:mudiff}) is the solution to the Laplace equation (\ref{eq:ohm}), fixing the current flow at $r=\infty$ and placing a hard obstacle of radius $\sqrt{B}$ in the way.   Hence, we identify $B=R_{\mathrm{eff}}^2$, with $R_{\mathrm{eff}}$ defined in (\ref{eq:main}).   

As we discussed previously, so long as the total sample size $L \gg \lambda$,   because only the diffusive sector in hydrodynamics persists on the length scale $L$,  we do not expect hydrodynamic effects near the boundary to qualitatively destroy (\ref{eq:main}).   In particular, the most important hydrodynamic contributions to `contact' resistances will be present with or without the obstacle present, and will be subtracted off in the measurement of $\mathcal{R}^\prime$.

\subsection{No-Stress Boundary Conditions}
Another reasonable choice of boundary conditions is no-stress boundary conditions.   As before, we demand that $v_r=0$ at $r=R$.   We additionally impose that $T_{r\theta} = 0$ -- this is the statement that no transverse momentum can flow into the boundary.   We will see a simple model for how these boundary conditions arise in kinetic theory in Section \ref{sec3}.   On general physical grounds, no-stress boundary conditions are a much more plausible assumption for electronic fluids than for classical fluids like water.   Hydrodynamics is the study of the densities and fluxes of conserved quantities, which are well-defined regardless of the `frame' of hydrodynamic variables (a non-trivial statement in inhomogeneous fluids \cite{landau}) -- so the most natural choice of boundary conditions for a quantum fluid are that there is no charge and transverse momentum flux through the obstacle.    There is experimental evidence that these are the correct boundary conditions in graphene \cite{bandurin, polini}.    Using \cite{landau} \begin{equation}
T_{r\theta} = -\eta \left(\frac{1}{r}\partial_\theta v_r + \partial_r v_\theta - \frac{v_\theta}{r}\right) = 0,
\end{equation}
we find a new boundary condition in terms of the stream function: \begin{equation}
\frac{1}{r}\partial_r \psi + \frac{1}{r^2} \partial_\theta^2 \psi - \partial_r^2 \psi  = 0.   \label{eq:18}
\end{equation}
The solution proceeds identically to before.  Starting from the general solution (\ref{eq:genhydro}),  boundary conditions at $r=\infty$ impose $A=u$ and $C=0$.    Imposing the boundary conditions (\ref{eq:vr0}) and (\ref{eq:18}), and applying more Bessel function identities, we find \begin{subequations}\label{eq:nostressfinal}\begin{align}
B &= -R^2 \dfrac{\displaystyle \frac{R^2}{\lambda^2}\mathrm{K}_3\left(\frac{R}{\lambda}\right) + \left(3\frac{R^2}{\lambda^2}+8\right)\mathrm{K}_1\left(\frac{R}{\lambda}\right) + 4\frac{R}{\lambda}\mathrm{K}_0\left(\frac{R}{\lambda}\right)}{\displaystyle \frac{R^2}{\lambda^2}\mathrm{K}_3\left(\frac{R}{\lambda}\right) + \left(3\frac{R^2}{\lambda^2}-8\right)\mathrm{K}_1\left(\frac{R}{\lambda}\right) + 4\frac{R}{\lambda}\mathrm{K}_0\left(\frac{R}{\lambda}\right)}  ,  \label{eq:lambda12} \\
D &= \dfrac{16R}{\displaystyle \frac{R^2}{\lambda^2}\mathrm{K}_3\left(\frac{R}{\lambda}\right) + \left(3\frac{R^2}{\lambda^2}-8\right)\mathrm{K}_1\left(\frac{R}{\lambda}\right) + 4\frac{R}{\lambda}\mathrm{K}_0\left(\frac{R}{\lambda}\right)}.
\end{align}\end{subequations}
Although the functional form of this answer is less elegant than before, remarkably we find that the asymptotic behavior of $\psi$ as $r\rightarrow\infty$ is identical to (\ref{eq:asy1}) when $R\gg \lambda$,  and to (\ref{eq:asy2}) when $R\ll \lambda$.   (\ref{eq:mudiff}) continues to hold.

\section{Kinetic Theory}\label{sec3}
Kinetic theory is an effective semiclassical description of quasiparticle dynamics on long length scales compared to the typical wavelength of the quasiparticles, $1/k_{\mathrm{F}}$ \cite{kamenev}.   Although we neglect quantum phase coherence, we do assume that the number density $f(\mathbf{x},\mathbf{p})$ of quasiparticles of each momentum $\mathbf{p}$ must be accounted for, at each point in space $\mathbf{x}$.   This number density is not strictly conserved, but as in our treatment of hydrodynamics with momentum relaxation, we imagine that the non-conserved modes of  $f(\mathbf{x},\mathbf{p})$ decays over a time scale `long enough' to keep track of.   For us, this occurs when the typical particle velocity $v_{\mathrm{F}}$ is fast enough that the decay rate of the non-conserved modes is slow compared to $v_{\mathrm{F}}/R$.    Assuming no external forces act on the electron gas, and assuming steady-state flow, the equations of motion of kinetic theory become \begin{equation}
\mathbf{v}(\mathbf{p})\cdot \partial_{\mathbf{x}} f(\mathbf{x},\mathbf{p}) = \mathcal{C}[f].  \label{eq:boltzmann}
\end{equation}
The left hand side of this equation describes motion of free quasiparticles along trajectories $\mathrm{d}\mathbf{x} / \mathrm{d}t = \mathbf{v}_\mathbf{p}$;  the right hand side calculates the scattering of these quasiparticles off of each other.    

In this paper, we make four further simplifications, following \cite{molenkamp, levitov1607}.    Firstly, we assume that there is a single, rotationally symmetric, Fermi surface of Fermi momentum $p_{\mathrm{F}}$ and Fermi velocity $v_{\mathrm{F}}$.   Secondly, we assume that thermal effects are negligible, and so all particles approximately stay ``exactly" on the Fermi surface:  their momenta all obey $|\mathbf{p}| = p_{\mathrm{F}}$.   Quasiparticles can be parameterized by the direction of their motion $\phi$:  \begin{equation}
\mathbf{v}_{\mathbf{p}} = v_{\mathrm{F}}\left(\cos\phi \hat{\mathbf{x}} + \sin\phi \hat{\mathbf{y}}\right),
\end{equation}
and the distribution function simplifies to $f(\mathbf{x},\phi)$.   Thirdly, we will assume a simple, linear form for the collision operator.   To define $\mathcal{C}$ more precisely, it is convenient to write \begin{equation}
f(\mathbf{x},\phi) =  \frac{1}{2\mpi}\sum_{m\in\mathbb{Z}} A_m(\mathbf{x}) \mathrm{e}^{\mathrm{i}m\phi}.
\end{equation}
We then define \begin{equation}
\mathcal{C}[f] \equiv - \frac{v_{\mathrm{F}}}{\xi}\sum_{|m|\ge 1} \frac{1}{2\mpi} A_m(\mathbf{x}) \mathrm{e}^{\mathrm{i}m\phi} - \frac{v_{\mathrm{F}}}{\ell_0 }\sum_{|m|\ge 2} \frac{1}{2\mpi} A_m(\mathbf{x}) \mathrm{e}^{\mathrm{i}m\phi}.  \label{eq:Cop}
\end{equation}
Electron-electron collisions occur at a rate $v_{\mathrm{F}}/\ell_0$, and they do not affect the $m=-1,0,1$ harmonics because, as we will see, these harmonics encode charge and momentum, which are conserved in electron-electron scattering.   The parameter $\xi$ is the length scale over which electron-impurity scattering occurs.  These collisions conserve only charge ($m=0$).   Combining (\ref{eq:mattheisen}), (\ref{eq:boltzmann}) and (\ref{eq:Cop}), and defining \begin{equation}
\mathcal{C}_m = \left\lbrace \begin{array}{ll} 0 &\  m =0 \\ \xi^{-1} &\ |m|=1 \\ \ell^{-1} &\ |m|\ge 2\end{array}\right.,  \label{eq:CalCM}
\end{equation}
 we obtain \begin{equation}
(\partial_x + \mathrm{i}\partial_y) A_{m+1}(\mathbf{x}) + (\partial_x - \mathrm{i}\partial_y) A_{m-1}(\mathbf{x})  = - 2\mathcal{C}_m A_m(\mathbf{x}).  \label{eq:boltzmann2}
\end{equation} 
Finally, as we discussed in Section \ref{sec2}, we will neglect long range Coulomb interactions, as we expect that (as in hydrodynamics) they can be absorbed into a (non-local) redefinition of the chemical potential, which we will see is related to $A_0$.

In two-dimensional materials such as graphene, in the quasiparticle limit scattering processes are likely to be dominated by collinear processes, and so there may be a richer angular structure to the collision terms than we have presented above.   This may lead to logarithmic corrections to the above theory \cite{logT2, polini1506}.    Furthermore, if the dominant contribution to the bulk resistivity involves scattering off of very long wavelength potential disorder, then the inhomogeneity should be explicitly accounted for in kinetic theory, instead of simply adding the scattering rate $1/\xi$ as we have done in (\ref{eq:CalCM}).   As this is a significantly more challenging computation, we will stick to the simple homogeneous model for disorder introduced above.  With these two caveats in mind, we expect that this simple kinetic approach will capture the qualitative physics accessed in experiment.

\subsection{The Hydrodynamic Limit}\label{sec:hydro3}
We begin by deriving hydrodynamics as a limit of (\ref{eq:boltzmann2}), in the simple kinetic theory model introduced above.   Consider a distribution function $f$ which is smoothly varying on length scales $\gtrsim d$.   We take $d\gg \ell$,  but $d$ may be large or small compared to $\xi$.   Then we expect that \emph{schematically}, (\ref{eq:boltzmann2}) becomes \begin{subequations}\begin{align}
\frac{1}{d} A_{1} + \frac{1}{d}A_{-1} &\sim 0 \\
\frac{1}{d}A_0+ \frac{1}{d}A_2 &\sim \frac{1}{\xi} A_1, \\
\frac{1}{d}A_1 + \frac{1}{d}A_3 &\sim \frac{1}{\ell} A_2,  \\
\frac{1}{d}A_2 + \frac{1}{d}A_4 &\sim \frac{1}{\ell}A_3,
\end{align}\end{subequations}
and so on.   These equations are consistent with the ansatz \begin{equation}
A_m \sim \left(\frac{l}{d}\right)^{m-1} A_1, \;\; (m\ge 1).  \label{eq:sec32am}
\end{equation}
Hence, it is natural to imagine ``throwing out" modes with $|m|>k$, with $k$ some ``critical" value.   Since we know that $d\gg \ell$, a natural choice is $k=2$ -- this is the first mode which is forcibly ``small" according to (\ref{eq:sec32am}).   More precisely, we approximate \begin{subequations}\begin{align}
(\partial_x - \mathrm{i}\partial_y) A_1 &\approx -\frac{2}{\ell}A_2, \\
(\partial_x + \mathrm{i}\partial_y) A_{-1} &\approx -\frac{2}{\ell}A_{-2}.
\end{align}\end{subequations}
(\ref{eq:boltzmann2}) can then be closed into a set of three coupled equations: \begin{subequations}\label{eq:A101}\begin{align}
(\partial_x - \mathrm{i}\partial_y)A_0 -\frac{\ell}{2}\left(\partial_x^2 + \partial_y^2\right) A_1 &= -\frac{2}{\xi}A_1, \\
(\partial_x + \mathrm{i}\partial_y)A_0 -\frac{\ell}{2}\left(\partial_x^2 + \partial_y^2\right) A_{-1} &= -\frac{2}{\xi}A_{-1}, \\
(\partial_x + \mathrm{i}\partial_y)A_1 + (\partial_x - \mathrm{i}\partial_y)A_{-1} &= 0.
\end{align}\end{subequations}

To make the connection with the hydrodynamic equations (\ref{eq:hydro1}) and (\ref{eq:hydro2}) precise, we now claim that the hydrodynamic pressure and velocity fields are related to kinetic theory parameters as: \begin{subequations}\label{eq:Psec32}\begin{align}
P &= \frac{mv_{\mathrm{F}}^2}{2} A_0, \\
v_x &= \frac{v_{\mathrm{F}}}{2n_0}  (A_1+A_{-1}), \\
v_y &= \frac{v_{\mathrm{F}}}{2n_0}  \mathrm{i}(A_1-A_{-1}), 
\end{align}\end{subequations}
with $m_0$ the quasiparticle mass.    The justification of (\ref{eq:Psec32}) is as follows.   Firstly, by definition \begin{equation}
A_0 = n,
\end{equation} 
the number density of quasiparticles.   In linear response, neglecting thermal fluctuations, we know that $P$ is a function of $n$.   In linear response \cite{landau},  \begin{equation}
P-P_0 \approx (n-n_0) m_0 c_{\mathrm{s}}^2 
\end{equation}where $c_{\mathrm{s}} = v_{\mathrm{F}}/\sqrt{2}$ is the speed of sound in this gas.   This result for $c_{\mathrm{s}}$ can be found linearizing the time-dependent generalization of (\ref{eq:boltzmann}) \cite{levitov1607}.   We can neglect the constant pressure $P_0$, as only derivatives of $P$ enter (\ref{eq:hydro2}).    The definition of the velocity follows from the fact that the particle current is given by \begin{subequations}\begin{align}
n_0 v_x(\mathbf{x}) &= \int \mathrm{d}\phi  \; (v_{\mathrm{F}}\cos\phi) f(\mathbf{x}, \phi) = v_{\mathrm{F}} \frac{A_1+A_{-1}}{2}, \\
n_0 v_y(\mathbf{x}) &= \int \mathrm{d}\phi  \; (v_{\mathrm{F}}\sin\phi) f(\mathbf{x}, \phi) = v_{\mathrm{F}} \frac{A_{-1}-A_{1}}{2\mathrm{i}}.
\end{align}\end{subequations}

Plugging (\ref{eq:Psec32}) into (\ref{eq:A101}) recovers (\ref{eq:hydro2}) from the first and second equations of (\ref{eq:A101}), and (\ref{eq:hydro1}) from the third.   The viscosity is given by \begin{equation}
\eta = \frac{m_0n_0v_{\mathrm{F}}\ell}{4} 
\end{equation}
and the momentum relaxation parameter is given by \begin{equation}
\Gamma = \frac{m_0n_0 v_{\mathrm{F}}}{\xi}.  \label{eq:Gammakin}
\end{equation}
Combining (\ref{eq:rho}) and (\ref{eq:Gammakin}) , we can recover the classic Drude formula for the resistivity (\ref{eq:Druderho}), as advertised in the introduction.   Another useful formula is for the momentum relaxation length in hydrodynamics, in terms of microscopic parameters.   Using (\ref{eq:lambda}), we find \begin{equation}
\lambda = \frac{\sqrt{\ell \xi}}{2}.  \label{eq:lambda2}
\end{equation}
These results are exactly analogous to those found in a somewhat more rigorous treatment in \cite{polini1506}.

\subsection{Estimating the Effective Scattering Length} 
Of course, our real aim is to solve the problem of flow around a cylinder beyond the hydrodynamic limit.   So we wish to analyze the kinetic equations on length scales short compared to $\ell$, when all of the modes $A_m$ need to be included.   In this subsection, we present a heuristic argument for the physics of hard cylinder scattering.   

Let us consider a set-up similar to \cite{levitov1607}.  Instead of a hard cylinder of radius $R$, we consider deforming the collision term in the Boltzmann equation (\ref{eq:boltzmann}) to \begin{equation}
\mathcal{C}_{\mathrm{new}}[f] = \mathcal{C}[f] - v_{\mathrm{F}} \alpha(\mathbf{x}) \sum_{m=\pm 1} \frac{1}{2\mpi} A_m(\mathbf{x}) \mathrm{e}^{\mathrm{i}m\phi},
\end{equation}
with $\alpha(\mathbf{x})$ taken (for now) to be perturbatively small,  and only non-zero in a region of radius $\sim R$.   At the end of the day, we want to take $\alpha \rightarrow \infty$ to mimic some kind of ``hard cylinder scattering".   The important thing is simply that scattering can only occur if $r\lesssim R$.  The fact that we have only chosen to include the $\pm 1$ modes in the $\alpha$ term is to make the resulting computation exactly solvable.   

We begin in the opposite limit of $\alpha \rightarrow \infty$, instead treating $\alpha$ as a perturbatively small parameter.  Suppose that we look for a solution to the modified Boltzmann equation of the form \begin{equation}
f(\mathbf{x},\phi) = f_0(\mathbf{x},\phi) + \hat f(\mathbf{x},\phi),
\end{equation}
with $f_0(\mathbf{x},\phi)$ corresponding to a constant velocity field, and $\hat f$ perturbatively small.   Closely mimicking \cite{levitov1607} in what follows,  we may write the linear Boltzmann equation in abstract form \begin{equation}
\mathsf{G}^{-1}(\mathbf{f}_0 + \hat{\mathbf{f}}) = \mathsf{B}(\mathbf{f}_0 + \hat{\mathbf{f}}),  \label{eq:linalg}
\end{equation}
where the (infinite) matrices \begin{subequations}\begin{align}
\mathsf{G}^{-1} &\equiv \cos\phi \partial_x + \sin \phi \partial_y + \frac{1}{\ell} - \frac{1}{\ell} \mathsf{Q} + \frac{1}{\xi} \mathsf{P}, \\
\mathsf{B} &\equiv  - \alpha(\mathbf{x}) \mathsf{P}
\end{align}\end{subequations} and we have defined projection matrices onto low Fourier modes:\begin{subequations}\begin{align}
\mathsf{Q} f &= \frac{1}{2\mpi} \sum_{m=-1}^1 A_m \mathrm{e}^{\mathrm{i}m\phi}, \\
\mathsf{P} f &= \frac{1}{2\mpi} \sum_{m=\pm 1} A_m \mathrm{e}^{\mathrm{i}m\phi}.
\end{align}\end{subequations}
Now, a formal solution to (\ref{eq:linalg}) is given by \begin{equation}
\hat{\mathbf{f}} = \left(\mathsf{G}^{-1}-\mathsf{B}\right)^{-1}\mathsf{B}  \mathbf{f}_0 = \left(\mathsf{GB} + \mathsf{GBGB} + \mathsf{GBGBGB} + \cdots\right)\mathbf{f}_0.
\end{equation}
Since we have assumed that $\mathsf{B}$ is perturbatively small, we will assume that the above Taylor series converges.   Now, let us focus on the low momentum modes, and simply worry about computing $\mathsf{Q}\hat{\mathbf{f}}$.  But now we notice that \begin{equation}
\mathsf{Q GB} \mathbf{f}_0 = -\alpha \mathsf{QGP} \mathbf{f}_0 = -\alpha \mathsf{QGQP} \mathbf{f}_0= \mathsf{QGQB} \mathbf{f}_0.
\end{equation}
To get this chain of equations, we have used basic properties of projection matrices, as well as the fact that $\mathsf{P}$ projects onto a subset of the non-projected vectors of $\mathsf{Q}$.   $\mathsf{QGQ}$ and $\mathsf{B}$ are only non-vanishing in the $m=-1,0,1$ directions.   With the identity $\mathsf{BGB} = \mathsf{BQGQB}$, which also follows from projector identities, we conclude that the 3-component vector $\mathsf{Q}\hat{\mathbf{f}} \equiv \hat{\mathbf{f}}^\prime$ can be found by solving the more tractable equation \begin{equation}
\hat{\mathbf{f}}^\prime = \left( \mathsf{G}^\prime \mathsf{B}  + \mathsf{G}^\prime \mathsf{B}\mathsf{G}^\prime \mathsf{B} + \cdots\right) \mathbf{f}_0^\prime .  \label{eq:gbg}
\end{equation}
with $\mathsf{G}^\prime \equiv \mathsf{QGQ}$.

Now, we need a useful expression for $\mathsf{G}^\prime$.    Since $\mathsf{G}^\prime$ does not have explicit position dependence, it will be useful to do a spatial Fourier transform.   To this end, let us write \begin{equation}
\mathsf{G}^{-1}(\mathbf{k}) = \mathsf{G}_0^{-1}(\mathbf{k}) - \mathsf{H},
\end{equation} with \begin{subequations}\begin{align}
\mathsf{G}_0^{-1}(\mathbf{k}) &\equiv \mathrm{i}k_x \cos \phi + \mathrm{i}k_y \sin \phi + \frac{1}{\ell}, \\
\mathsf{H} &\equiv \frac{1}{\ell} \mathsf{Q} - \frac{1}{\xi} \mathsf{P}.  \label{eq:sfh}
\end{align}\end{subequations}
Just as before, we can write \begin{equation}
\mathsf{G} = \mathsf{G}_0 + \mathsf{G}_0 \mathsf{HG}_0 + \cdots,  \label{eq:GGO}
\end{equation}
and as $\mathsf{QHQ} = \mathsf{H}$, by left and right multiplying (\ref{eq:GGO}) by $\mathsf{Q}$, and resumming the resulting Taylor series, we find \begin{equation}
\mathsf{G}^\prime = \mathsf{G}_0^\prime \left(1 - \mathsf{HG}_0^\prime\right)^{-1}  \label{eq:G0p2}
\end{equation}
with $\mathsf{G}_0^\prime \equiv \mathsf{QG}_0\mathsf{Q}$.   The components of $\mathsf{G}_0$ can be found explicitly.  In particular, in a basis of $\phi$-Fourier modes:  \begin{equation}
\left(\mathsf{G}_0(\mathbf{k})\right)_{mn} = \int\limits_0^{2\mpi}\frac{\mathrm{d}\phi}{2\mpi} \frac{\mathrm{e}^{\mathrm{i}(n-m)\phi} }{\ell^{-1} + \mathrm{i}k_x \cos\phi + \mathrm{i}k_y \sin \phi } = \int\limits_0^{2\mpi}\frac{\mathrm{d}\phi}{2\mpi} \frac{\mathrm{e}^{\mathrm{i}(n-m)(\phi^\prime + \phi_k)} }{\ell^{-1} + \mathrm{i} k \cos \phi^\prime },
\end{equation}
where $\tan \phi_k = k_y/k_x$.    Without loss of generality, we take $n-m\ge 0$.  For the remaining case, we may simultaneously complex conjugate and send $\mathbf{k} \rightarrow -\mathbf{k}$.   We may evaluate this integral through contour integration.   The integral over $\phi$ is equivalent to an integral over a complex variable $z$ on the unit circle $|z|=1$, which is evaluated using Cauchy's theorem:  \begin{align}
\left(\mathsf{G}_0(\mathbf{k})\right)_{mn} &= \frac{1}{2\mpi \mathrm{i}} \oint \mathrm{d}z \frac{z^{n-m}}{\ell^{-1}z + \mathrm{i}k (1+z^2)} \mathrm{e}^{\mathrm{i}(n-m)\phi_k} =  \frac{\mathrm{e}^{\mathrm{i}(n-m)\phi_k} (-\mathrm{i})^{n-m}}{\sqrt{k^2 + \ell^{-2}}} \left(\sqrt{1+\frac{1}{(k\ell)^2}} -\frac{1}{k\ell}\right)^{n-m} .
\end{align}
Generalizing to $n<m$ by the rules described earlier, we find \begin{equation}
\mathsf{G}_0^\prime = \frac{-1}{\sqrt{k^2+\ell^{-2}}} \left(\begin{array}{ccc} -1 &\ \mathrm{i}\mathrm{e}^{\mathrm{i}\phi_k} \left(\sqrt{1+\frac{1}{(k\ell)^2}} -\frac{1}{k\ell}\right) &\ \mathrm{e}^{2\mathrm{i}\phi_k} \left(\sqrt{1+\frac{1}{(k\ell)^2}} -\frac{1}{k\ell}\right)^2  \\  \mathrm{i}\mathrm{e}^{-\mathrm{i}\phi_k} \left(\sqrt{1+\frac{1}{(k\ell)^2}} -\frac{1}{k\ell}\right) &\ -1 &\  \mathrm{i}\mathrm{e}^{\mathrm{i}\phi_k} \left(\sqrt{1+\frac{1}{(k\ell)^2}} -\frac{1}{k\ell}\right) \\ \mathrm{e}^{-2\mathrm{i}\phi_k} \left(\sqrt{1+\frac{1}{(k\ell)^2}} -\frac{1}{k\ell}\right)^2 &\  \mathrm{i}\mathrm{e}^{-\mathrm{i}\phi_k} \left(\sqrt{1+\frac{1}{(k\ell)^2}} -\frac{1}{k\ell}\right) &\ -1 \end{array}\right).  \label{eq:G0P}
\end{equation}
Combining (\ref{eq:sfh}), (\ref{eq:G0p2}) and (\ref{eq:G0P}) we can compute $\mathsf{G}^\prime$.   As we will see soon, we are going to want to re-sum the series (\ref{eq:gbg}), and so let us merely compute   \begin{equation}
\mathsf{PG}(\mathbf{k})\mathsf{P} = \frac{\ell \xi}{ \xi \sqrt{(k\ell)^2+1} - \xi + 2\ell } \left(\begin{array}{cc} 1 &\ -\mathrm{e}^{2\mathrm{i}\phi_k} \\ -\mathrm{e}^{-2\mathrm{i}\phi_k} &\ 1 \end{array}\right).  \label{eq:PGP}
\end{equation}

Now, let us return to (\ref{eq:gbg}).  In Fourier space,  $\alpha(\mathbf{x})$ looks \emph{heuristically} like \begin{equation}
\alpha(\mathbf{k}, \mathbf{k}^\prime) \sim \alpha_0 \mathrm{\Theta}\left(|\mathbf{k}-\mathbf{k}^\prime|-\frac{1}{R_*}\right)
\end{equation}
with $\alpha_0$ a perturbatively small parameter, $\mathrm{\Theta}$ the Heaviside step function, and $R_*\sim R$ a cut-off at scales of order the size of the cylinder.    Let us make the further assumption that $\mathbf{k}$ and $\mathrm{k}^\prime$ are also separately constrained to be approximately $\lesssim R_*^{-1}$.   In this case, we approximate that \begin{equation}
\mathsf{B} \mathsf{G}^\prime \mathsf{B} \sim g \mathsf{B}^2 ,  \label{eq:Gprimeend}
\end{equation}  with \begin{equation}
g \equiv  \int\limits \frac{\mathrm{d}^2\mathbf{k}}{(2\mpi)^2}  \mathrm{\Theta}\left(k-\frac{1}{R_*}\right) \frac{\ell}{\sqrt{(k\ell)^2 + 1} -1 + 2\ell \xi^{-1}} .   \label{eq:gint}
\end{equation}
(\ref{eq:Gprimeend}) comes from integrating the kernel $\mathsf{G}^\prime$ over long wavelengths.   The off-diagonal components of $\mathsf{PGP}$ vanish because of the complex angular factors (\ref{eq:PGP}).   Integrating over angles in (\ref{eq:gint}), and switching variables to $u = \sqrt{(k\ell)^2+1} -1 + 2\ell \xi^{-1}$, we obtain \begin{align}
g &= \frac{1}{2\mpi \ell} \int\limits_{2\ell/\xi}^{\sqrt{1+(\ell/R_*)^2} - 1 + 2\ell/\xi} \frac{\mathrm{d}u}{u} \left(u+1-\frac{2\ell}{\xi}\right) \notag \\
&= \frac{1}{2\mpi \ell} \left[\left(1-\frac{2\ell}{\xi}\right)\log \left(\frac{\xi}{2\ell}\left(\sqrt{1+\left(\frac{\ell}{R_*}\right)^2}-1\right) + 1\right) + \sqrt{1+\left(\frac{\ell}{R_*}\right)^2}-1\right].   \label{eq:geq}
\end{align}

We are now able to heuristically sum the series (\ref{eq:gbg}).   What we find is that, as in \cite{levitov1607},  $\alpha_0$ is `renormalized' to \begin{equation}
\alpha^\prime = \alpha_0  - \alpha_0^2 g + \alpha_0^3 g^2 + \cdots = \frac{\alpha_0}{1+\alpha_0 g}.
\end{equation}
Now, we return to the problem we are actually studying -- scattering off of a hard cylinder.  The hard cylinder does not allow any fluid to flow through it, so we expect $\alpha_0 \rightarrow \infty$.   Then we find that \begin{equation}
\alpha^\prime = \frac{1}{g}.
\end{equation}
Without explicitly solving (\ref{eq:gbg}) for $\hat{\mathbf{f}}$, we do know that it is approximately proportional to $\alpha^\prime$.    In our hydrodynamic model, we found that the asymptotic response of the chemical potential was $\mu \sim 1/(\xi r)$ as $r\rightarrow \infty$ (the factor of $\xi$ comes from the resistivity $\rho$), and we will confirm the same thing numerically in kinetic theory.   So we conclude that $\mu \sim 1/(\xi gr)$, and following the logic of (\ref{eq:Reff}), we postulate  \begin{equation}
R_{\mathrm{eff}}^2 \equiv \frac{C\xi }{ g},  \label{eq:Rg}
\end{equation}
with $C$ an O(1) constant.   We fix $C$ and $R_*$ by now demanding that (\ref{eq:Rg}) asymptotically match analytic results in the hydrodynamic regime.

(\ref{eq:geq}) suggests that $R_{\mathrm{eff}}$ should not depend much on the boundary conditions at the surface of the cylinder.   Within hydrodynamics, we have already seen this effect.    By taking $\ell \rightarrow 0$ with fixed $\lambda$, we may compare (\ref{eq:geq}) and (\ref{eq:Rg}) to our exact asymptotic hydrodynamic results (\ref{eq:asy1}) and (\ref{eq:asy2}).  Demanding that (\ref{eq:Rg}) matches (\ref{eq:asy1}) requires that when $\ell \rightarrow 0$, $\ell \ll \xi$ and $\sqrt{\ell \xi} \ll R$: \begin{equation}
 R^2 = C\xi \left(\frac{\xi}{8\mpi R_*^2}\right)^{-1} = 8\mpi CR_*^2.
\end{equation} 
Employing (\ref{eq:lambda2}) and  demanding that (\ref{eq:Rg}) matches (\ref{eq:asy2}) when $\lambda \gg R$:  \begin{equation}
\frac{2\lambda^2}{\log\frac{\lambda}{R}} = C\xi \left(\frac{1}{2\mpi \ell} \log \frac{\ell\xi}{4R_*^2}\right)^{-1} = 4\mpi C \frac{\lambda^2}{\log \frac{\lambda}{R_*}}.
\end{equation}
These equations imply \begin{subequations}\label{eq:CRstar}\begin{align}
C &= \frac{1}{2\mpi}, \\
R_* &= \frac{R}{2}.
\end{align}\end{subequations}
Combining (\ref{eq:geq}), (\ref{eq:Rg}) and (\ref{eq:CRstar}) we obtain (\ref{eq:main}).

\begin{figure}
\centering
\includegraphics{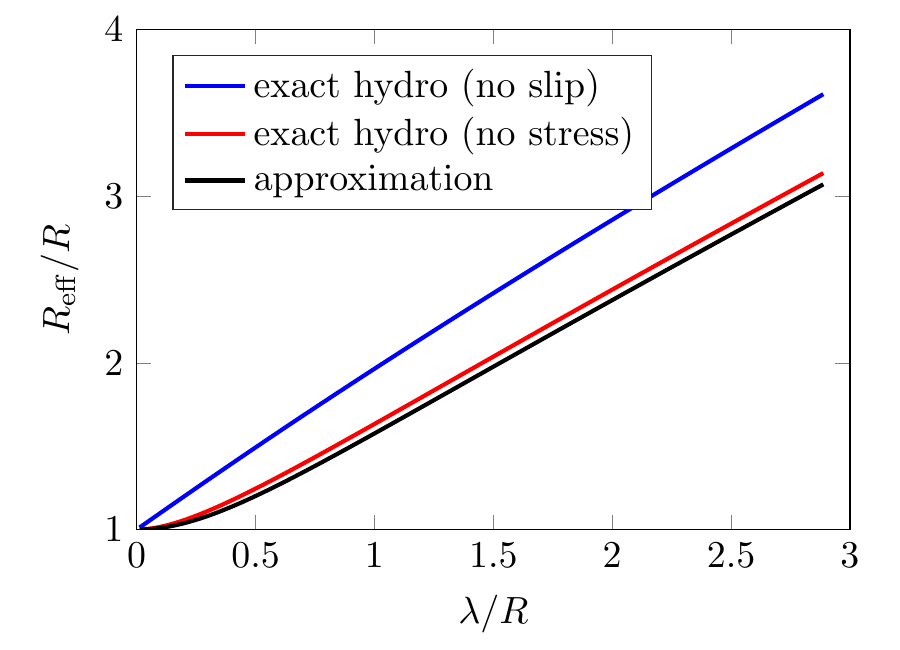}
\caption{A comparison of our approximate result (\ref{eq:main}), when $\ell \ll R$, to the exact hydrodynamic results (\ref{eq:noslipfinal}) for no slip boundary conditions, and (\ref{eq:nostressfinal}) for no stress boundary conditions.   While all three curves will agree both as $\lambda/R \rightarrow 0$ and $\lambda/R\rightarrow \infty$, the latter occurs logarithmically slowly.}
\label{fig:approxhydro}
\end{figure}

As shown in Figure \ref{fig:approxhydro}, when $\ell \ll R$ our simple approximation is embarassingly good for any ratio of $R/\lambda$.    In fact, the worst discrepancies are $\lesssim 25\%$ for no slip boundary conditions, and  $\lesssim 5\%$ for no stress boundary conditions.  So we expect that it may capture the physics of the ballistic scattering regime $R\ll \ell$ reasonably well too.  

\subsection{Rotational Invariance}
We now wish to confirm our heuristic argument above with a numerical solution to the Boltzmann equation.  Before putting (\ref{eq:boltzmann}) on a computer, we will first show, in this subsection, how to take advantage of rotational symmetry in order to reduce the number of equations we must solve.   In the next subsection, we will describe a simple and physically plausible set of boundary conditions.

Rotational invariance is implemented in (\ref{eq:boltzmann}) in a slightly subtle way.   If we rotate the system by a global angle $\theta^\prime$, then in cylindrical coordinates we send $(r,\theta) \rightarrow (r,\theta+\theta^\prime)$.   But we also rotate the quasiparticle momentum, and so $\phi \rightarrow \phi + \theta^\prime$.  Hence, we define a rotationally invariant angle \begin{equation}
\omega \equiv \phi - \theta,
\end{equation}and write \begin{equation}
f(r,\theta,\phi) = \frac{1}{2\mpi} \sum_{m,n\in \mathbb{Z}} a_m^n(r) \mathrm{e}^{\mathrm{i}m\omega + \mathrm{i}n\theta}.  \label{eq:amn}
\end{equation}

We claim that (\ref{eq:boltzmann}) will now look simple in the variables $a_m^n$.   First, let us naively do a coordinate change to the streaming terms:  \begin{equation}
\frac{1}{v_{\mathrm{F}}}\mathbf{v}\cdot\partial_{\mathbf{x}} = \cos\phi \left(\cos\theta \partial_r - \frac{\sin\phi}{r}\partial_\theta\right) + \sin\phi \left(\sin \theta \partial_r + \frac{\cos\phi}{r}\partial_\theta\right) = \cos \omega \partial_r + \frac{\sin\omega}{r}\partial_\theta.  \label{eq:73}
\end{equation}
But this is not quite what we want:  $\partial_\theta$ was defined above at fixed $\phi$, but we actually wish to take this derivative at fixed $\omega$.   We can think of this as follows: we write $f(\theta,\phi)$ as $\tilde f(\theta,\omega) =  f(\theta, \phi-\theta)$;  hence \begin{equation}
\partial_\theta  f(\theta,\phi) = (\partial_\theta - \partial_\omega) \tilde f(\theta,\theta+\omega).
\end{equation}
So in fact, we must replace (\ref{eq:73}) with \begin{equation}
\frac{1}{v_{\mathrm{F}}}\mathbf{v}\cdot\partial_{\mathbf{x}} =\cos \omega \partial_r + \frac{\sin\omega}{r} (\partial_\theta - \partial_\omega),  \label{eq:75}
\end{equation}
with derivatives now taken in the standard way on functions of the variables $r$, $\theta$ and $\omega$.   (\ref{eq:75}) does not depend on $\theta$ explicitly.   Hence, we conclude that the resulting equations are manifestly rotationally symmetric, and so the distinct Fourier modes, labeled by $n$, will decouple.   We can explicitly see this by plugging (\ref{eq:amn}) into (\ref{eq:boltzmann}), and employing (\ref{eq:75}) for the streaming terms.  The result is \begin{equation}
\left(\partial_r + \frac{n-(m-1)}{r}\right)a_{m-1}^n + \left(\partial_r - \frac{n-(m+1)}{r}\right)a^n_{m+1} = -2\mathcal{C}_m a_m^n  \label{eq:boltznum}
\end{equation}
with $\mathcal{C}_m$ defined in (\ref{eq:CalCM}).   As advertised, modes of distinct $n$ have decoupled.   As in the hydrodynamic problems of Section \ref{sec2}, the flow around a cylinder will excite only $n=\pm 1$ modes.

\subsection{Boundary Conditions}
In order to solve these equations (especially numerically) we must correctly identify the number of boundary conditions to impose.   As in Section \ref{sec:hydro3}, let us imagine truncating the equations to modes with $|m| \le k$; we have just seen that we may keep $n$ fixed by rotational symmetry.  Now we can concretely ask how many boundary conditions are required.   A hint comes from considering the case $k=2$, where we have seen how the kinetic equations reduce to hydrodynamics identically.   The hydrodynamic problem has 4 boundary conditions, as the stream function obeys a fourth order differential equation.   In kinetic theory, there are 5 variables when $k=2$:  from (\ref{eq:boltznum}) it appears that each obeys a first order differential equation.   The resolution to this puzzle is that there is a special combination of (\ref{eq:boltznum}) which has no radial derivatives (i.e., it is a constraint).  Consider the combination \begin{equation}
-2\mathcal{C}_k a_k^n + 2\mathcal{C}_{k-2} a_{k-2}^n - \cdots \pm 2\mathcal{C}_{-k}a_{-k}^n = \partial_r \left(a^n_{k-1} - \left(a^n_{k-1} + a^n_{k-3}\right)+ \left(a^n_{k-3} + a^n_{k-5}\right) \cdots \pm a^n_{1-k}\right) +   \mathrm{O}\left(\frac{1}{r}\right).
\end{equation}
The last sign in the above equation depends on the number $k$.   We can see that each radial derivative on the right hand side will cancel in this telescoping sum.    So indeed, we find a set of differential equations for $2k+1$ variables, with $2k$ independent radial derivatives.

\begin{figure}
\centering
\includegraphics{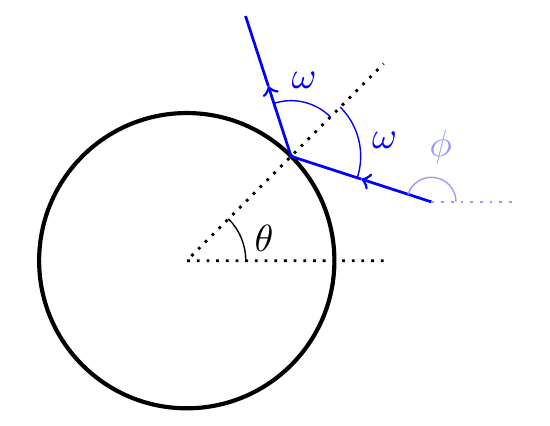}
\caption{A quasiparticle ballistically scattering off of a cylindrical obstacle.   We show the relationship between the angles $\theta$, $\phi$ and $\omega$.}
\label{fig:specular}
\end{figure}

Next, we think about what these boundary conditions must be.  We first think about the boundary conditions imposed by scattering off of the cylinder at $r=R$.   For simplicity, we will consider in this paper ``perfect" reflection:  namely, if the boundary is locally $x=0$,  an incoming particle with momentum $(-p_x,p_y)$ has outgoing momentum $(p_x,p_y)$.   As shown in Figure \ref{fig:specular}, this simply corresponds to $ \omega \rightarrow \mpi -\omega$.  Since $f$ counts the number density of quasiparticles, given the scattering process above, we conclude that \begin{equation}
f(r=R, \theta, \omega) = f(r=R,\theta, \mpi - \omega).
\end{equation}
Plugging this boundary condition into (\ref{eq:amn}), we see that \begin{equation}
a^n_m(R) = (-1)^m a^n_{-m}(R).  \label{eq:Rbc}
\end{equation}
If we only include $2k+1$ modes, since this boundary condition is trivial at $m=0$,  we conclude that there are $k$ boundary conditions at $r=R$.

We impose the remaining $k$ boundary conditions at $r=\infty$.  Firstly, from (\ref{eq:Psec32}) we conclude that (if the velocity at $r=\infty$ is imposed in the $x$-direction) we will need to impose boundary conditions that $a_{\pm 1}^{\pm 1}$ are constants.   The precise combination can be understood by re-writing (\ref{eq:amn}) for these modes in terms of $\theta$ and $\phi$:  \begin{equation}
f(\infty, \theta, \phi) = a^1_1 \mathrm{e}^{\mathrm{i}\phi} + a^{-1}_{-1} \mathrm{e}^{-\mathrm{i}\phi} + a^1_{-1} \mathrm{e}^{\mathrm{i}\phi - 2\mathrm{i}\theta} + a^{-1}_1 \mathrm{e}^{-\mathrm{i}\phi + 2 \mathrm{i}\theta} + \cdots 
\end{equation}
Since the velocity field is the same at all angles $\theta$, and $f$ should be real,  we conclude that the non-vanishing modes at $\infty$ are \begin{equation}
a^1_1(\infty) = a^{-1}_{-1}(\infty) = \frac{n_0 u}{v_{\mathrm{F}}}.
\end{equation}
The normalization factor comes from (\ref{eq:Psec32}).   It will actually be simpler to choose the boundary conditions \begin{equation}
a^1_1(\infty) = \frac{2n_0 u}{v_{\mathrm{F}}}.  \label{eq:a11}
\end{equation}
Since the differential equations are linear, we may simply take the real part of $f$ at the end.   (\ref{eq:a11}) is one more boundary condition.   We simply take the remaining $k-1$ boundary conditions to be \begin{equation}
a^1_m(\infty) = 0, \;\;\;\; (m=2,\ldots,k)
\end{equation}
and only solve for $n=1$ modes.   This ensures that $f$ asymptotes to uniform flow as $r\rightarrow \infty$.

There is one final technical point worth making.  A very simple exact solution to (\ref{eq:boltznum}), corresponding to a uniform flow, is \begin{equation}
a^1_m = \bar a_m \equiv  \frac{2n_0 u}{v_{\mathrm{F}}} \times  \left\lbrace \begin{array}{ll} -r/\xi &\ m=0 \\  1 &\ m=1 \\ 0 &\ m\ne 0,1  \end{array}\right..   \label{eq:abar}
\end{equation}
We show this more explicitly in Appendix \ref{app:diffkin}.   So in the numerics, we will write \begin{equation}
a^1_m = \bar a_m + \hat a_m,  \label{eq:hatam}
\end{equation}
and solve for $\hat a_m$, which is now finite in the entire domain of integration.

The boundary conditions we have discussed above are experimentally plausible in any material where electron scattering off of edges is very clean.   Recent experiments have demonstrated such nearly perfect ``specular" reflections in graphene \cite{dgg}.    These boundary conditions are analogous to the hydrodynamic no-stress boundary conditions.   The simplest way to see this is for each particle of velocity $\mathbf{v}$ 'entering' the obstacle, there is an equivalent one leaving, with $v_r$ reversed and $v_\theta$ unchanged.   Hence, the global fluid velocity $\sum v_r=0$ at the boundary, while there is no exchange of $\theta$-momentum across the boundary.   On the other hand, in other materials it has been argued that hydrodynamic flow appears more consistent with randomizing scattering, where momentum can be transferred into the obstacle \cite{molenkamp, mackenzie}.   We expect that the details of the boundary conditions do not qualitatively change our main result.

\subsection{Numerical Results}
We now solve the equations (\ref{eq:boltznum}) numerically for $\hat a_m$, as defined in (\ref{eq:hatam}), only including modes with $|m| \le k$.     By defining $u\equiv 1/r$, we can map our ``infinite" domain $r\ge R$ to the finite domain $u\in [0,1/R]$.  We use pseudospectral methods \cite{trefethen} to discretize the resulting differential equations for $\hat a_m(u)$.   Despite the presence of an essential singularity for most modes $\hat a_m(u)$ at $u=0$ (the exponential falloffs in the Bessel functions $\mathrm{K}_1(r/\lambda) \sim \mathrm{e}^{-r/\lambda}$ in Section \ref{sec2} become $\mathrm{e}^{-1/(u\lambda)}$), we find that these spectral methods are quite accurate.  They are also very computationally efficient, requiring runtimes of under a minute, even for $k\sim 50$.   Appendix \ref{app:conv} presents details on the rate at which our numerical methods converge.   

\begin{figure}
\centering
\includegraphics{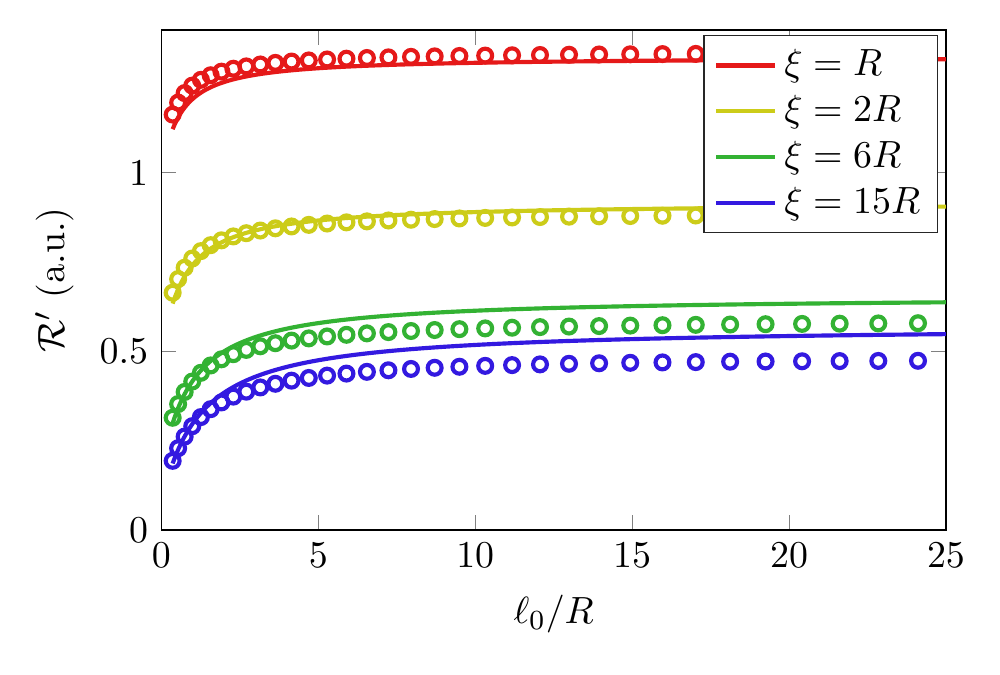}
\caption{A comparison of $\mathcal{R}^\prime$, as `predicted' by (\ref{eq:main}) and as computed numerically.   $\mathcal{R}^\prime$ decreases through either increasing $\xi$ or decreasing $\ell$, as predicted in  (\ref{eq:main}).  The maximal discrepancy between the heuristic analytic result and the numerical results is about 20\%.  As estimated in Appendix \ref{app:conv}, numerical error is not larger to the marker size.   }
\label{fig:data}
\end{figure}

Figure \ref{fig:data} shows a comparison of our numerical computations of $\mathcal{R}^\prime$ to our heuristic theoretical result (\ref{eq:main}).    The most important result is that there is no qualitative disagreement between (\ref{eq:main}) and the numerical computation which implements the proper boundary conditions.      In fact, the discrepancies shown in Figure \ref{fig:data} are no larger than 20\%, which is remarkable given the relative lack of rigor in our ``derivation" of (\ref{eq:main}) earlier.   The saturation of $\mathcal{R}^\prime$ as $\ell_0$ increases is simply a consequence of the fact that $\ell \approx \xi$ once $\ell_0 \gg \xi$.

\begin{figure}
\centering
\includegraphics[width=6.7in]{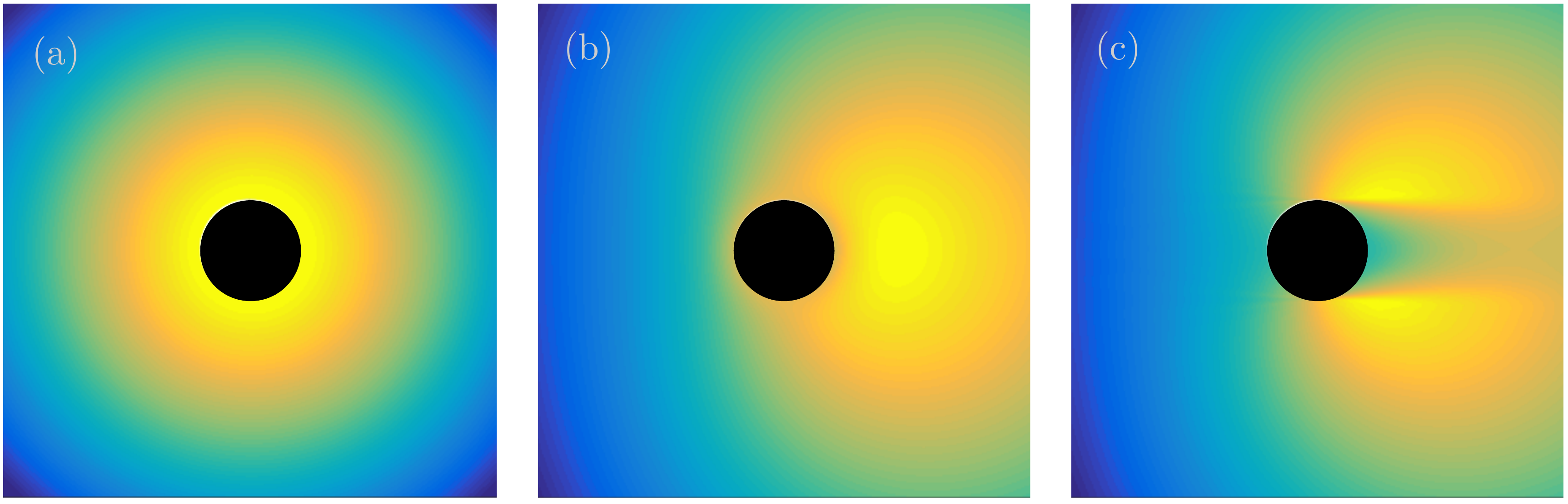}
\caption{Polar plots of $f(r,\omega)$, evaluated at $\theta=0$.   Lighter/yellower colors are larger values of $f$, and darker/bluer values are smaller.  (a) diffusive limit:  $\xi=\ell = 0.05R$.   The distribution of quasiparticles is completely dependent of internal velocity;  the density has radial dependence proportional to the chemical potential.  (b) hydrodynamic limit: $\xi = 4R$, $\ell = 0.05R$.  $f$ is dominated by the $|m|\le 1$ modes, but close to $r\sim R$ the distribution is approximately a sum of the constant velocity background than the diffusive background.  (c) ballistic limit: $\xi = 4R$,  $\ell = 1.5R$.  We see the `shadow' caused by the obstacle, as explained in the main text.  On scales $r\sim \ell$, the `shadow' fades away into the hydrodynamic background.}
\label{fig:scatter}
\end{figure}

It is also instructive to see the changes in $f(r,\theta,\omega)$ as $\ell$ becomes comparable to $R$.   Using angular symmetry, let us focus on $f(r,\theta=0,\omega) = f(r,\theta=0,\phi)$.   Using simple geometric considerations, we conclude that if $r/R < |\tan \omega|$, for $\omega\sim 0$, then a particle in this position must scatter off of the obstacle.   At $\theta=0$, the incoming particles will be at $\phi \sim \mpi$, and as $f\sim \cos \phi$ these particles are less occupied than average.   So we expect that within the fan $R\lesssim r \sin \omega$, there will be a `shadow' of depletion in $f(r,\omega)$.    Figure \ref{fig:scatter} confirms this prediction numerically.   The  observation of the `shadow' of the obstacle in the ballistic limit in our numerical simulations is a good qualitative consistency check.

The numerics will not be effective if $k\lesssim \ell /R$.   This limits the extent to which we may probe the ballistic limit.   We were able to comfortably simulate $\ell \lesssim 10R$, as shown in Figure \ref{fig:data}, and did not see any discrepancy with (\ref{eq:main}).    In the limit $\ell \lesssim 0.1 R$, our numerical results, using the kinetic theory code, agree with the hydrodynamic prediction (\ref{eq:lambda12}) to excellent precision of 5 or more digits.

%


%

\section{Experimental Outlook}\label{sec4}

So far, we have demonstrated the resolution to the Stokes paradox in momentum relaxing hydrodynamics, and in kinetic theory.   We conclude this paper by discussing the extent to which (\ref{eq:main}) can be observed in present day experiments.    We focus on the possibility of discovering this `Stokes paradox' in the Fermi liquid in doped graphene.   This has been experimentally observed at an electron density of about $n_0 \sim 10^{16} \; \mathrm{m}^{-2}$, at a temperature $T\sim 100$ K; this corresponds to the ratio $e\mu/k_{\mathrm{B}}T \sim 10$ \cite{bandurin}.  As such, we expect that thermal fluctuations can be neglected, and so our hydrodynamic and kinetic approximations are sensible.   More specifically, we can estimate the electron-impurity mean free path to be $\xi\sim 2$ $\mmu$m from resistivity data \cite{bandurin}, while one expects $\ell_0 \sim (150 \; \mathrm{K})^2/T^2 \times  0.4 $ $\mmu$m \cite{bandurin, polini1506}.\footnote{As noted previously,  the value of $\ell$ will pick up logarithmic corrections, but $\log(T/T_{\mathrm{F}}) \sim \log 10$ is not a large prefactor.}   

At least in graphene, the cylindrical obstacles in the Fermi liquid would likely be created through localized electrostatic probes, creating localized regions where chemical potential is substantially larger.  This can be done on scales as small as $R\sim 0.05$ $\mmu$m \cite{stroscio}.    Given that in Figure \ref{fig:FL1}, we saw the sharpest temperature decrease of $\mathcal{R}^\prime$ -- our key signature of hydrodynamics -- when $\ell \gtrsim R$,  this suggests that the observation of the phenomena predicted in this paper are within present day experimental reach.     

By tuning the temperature substantially above $100$ K, we expect electron-phonon scattering to dominate and for the Fermi liquid to behave diffusively.   Similarly, in the low temperature limit, electron-electron interactions are negligible and scattering off of the obstacle will be ballistic.    At temperatures of order 100 K, and for obstacles of (effective) size comparable to 0.4 $\mmu$m, we expect to see strong signatures of hydrodynamics.  Hence, measuring $\mathcal{R}^\prime$ may lead to an observation of all three regimes of behavior mentioned in the introduction in the same experimental setup.


%

\addcontentsline{toc}{section}{Acknowledgements}
\section*{Acknowledgements}
I would like to thank Arthur Barnard and Steven Kivelson for helpful discussions.   I was supported by the Gordon and Betty Moore Foundation.

\begin{appendix}

\section{The Diffusive Limit in Kinetic Theory}\label{app:diffkin}
In this appendix, we analytically solve (\ref{eq:boltznum}) for $n=1$ modes, assuming that $k=1$.  As we will see, we may solve the equations analytically, and the resulting solution corresponds to diffusive flow.   Such a solution is valid when $R\ll \xi$.   The equations we wish to solve read \begin{subequations}\begin{align}
-\frac{2}{\xi} a^1_{-1} &= \left(\partial_r - \frac{1}{r}\right)a^1_0,  \label{eq:appa1} \\
0 &= \left(\partial_r + \frac{2}{r}\right)a^1_{-1} + \partial_r a^1_1, \label{eq:appa2} \\
-\frac{2}{\xi} a^1_1 &= \left(\partial_r + \frac{1}{r}\right)a^1_0.   \label{eq:appa3}
\end{align}\end{subequations}
We make the ansatz that \begin{subequations}\begin{align}
a^1_1 &= \frac{2n_0 u}{v_{\mathrm{F}}}, \\
a^1_0 &= \frac{2n_0 u}{v_{\mathrm{F}}} \left[ c_1 r  + \frac{c_2}{r}\right], \\
a^1_{-1} &= \frac{2n_0 u}{v_{\mathrm{F}}}\left[ \frac{c_3}{r^2}\right].
\end{align}\end{subequations}
The boundary condition at $r=R$, (\ref{eq:Rbc}), implies that \begin{equation}
c_3 = -R^2.
\end{equation}
(\ref{eq:appa1}) then fixes \begin{equation}
c_2 = -\frac{R^2}{\xi},
\end{equation}
while (\ref{eq:appa3}) fixes \begin{equation}
c_1 = -\frac{1}{\xi}.
\end{equation}
(\ref{eq:appa2}) is satisfied by construction.  $a_0^1$ is exactly proportional to the pressure/chemical potential, as expected from (\ref{eq:mudiff}) and (\ref{eq:Psec32}).

If we take $R=0$, then we have shown (\ref{eq:abar}) is an exact solution to the $|m|\le 1$ equations.  In order to show that (\ref{eq:abar}) is an exact solution to the equations of motion for any $k$, we note that the only other equation of (\ref{eq:boltznum}) we need to consider is \begin{equation}
\partial_r a^1_1 + \left(\partial_r + \frac{2}{r}\right)a^1_3 = -\frac{2}{\ell} a^1_2.
\end{equation}
This is clearly solved by $a^1_1=\text{constant}$, with the other two modes vanishing.   

\section{Convergence of Numerical Methods}\label{app:conv}

\begin{figure}
\centering
\includegraphics{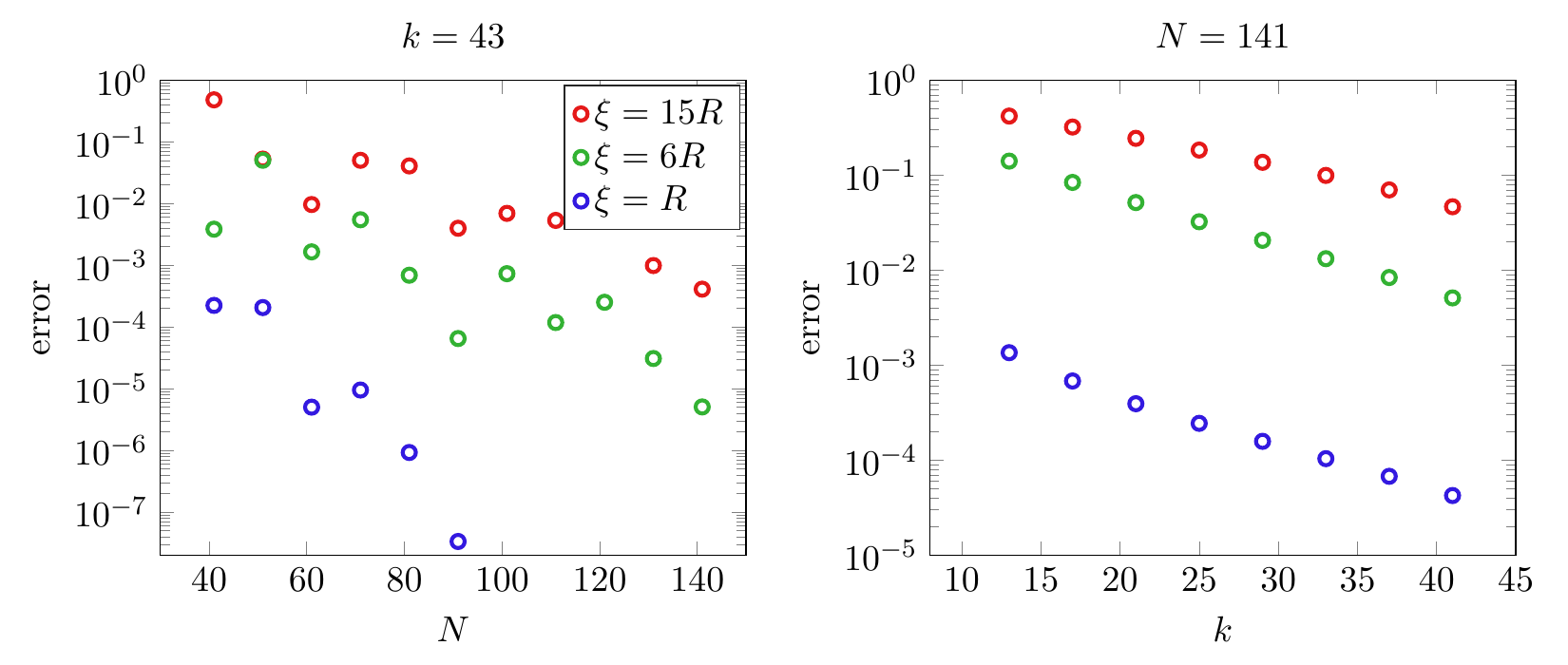}
\caption{Left panel:  we fix $k$ and vary $41 \le N \le 141$, using $N_*=181$ and $k_*=43$.  Right panel: we fix $N$ and vary $13\le k \le 41$, using $k_*=53$ and $N_*=141$.  Error is defined through (\ref{eq:error}).}
\label{fig:conv}
\end{figure}

There are two senses in which we require convergence for our numerical method:  in the number of grid points $N$ we use in the $u$ domain, and in the number of harmonics $k$ which we retain.   We measure the rate of convergence by defining the error in our numerical determination of $\mathcal{R}^\prime$ to be \begin{equation}
\text{error}_{N,k}(\mathcal{R}^\prime) \equiv \left| \frac{\mathcal{R}^\prime_{N,k}}{\mathcal{R}^\prime_{N_*,k_*}} -1 \right|  \label{eq:error}
\end{equation}  
Here $N_*$ and $k_*$ are reference values which should be `large'.    So long as $N$ and $k$ are not too close to $N_*$ and $k_*$, if the numerics are converging we expect the error to decrease exponential as we increase $N$ and $k$.
As expected for spectral methods, we see in Figure \ref{fig:conv} that there is exponential convergence in both $N$ and $k$ (despite more noise in the convergence in $N$).    All figures in the main text employ $N=181$ and $k=53$.   As we do not display data for $\xi>15R$ in this paper, based on simple extrapolations of the data in Figure \ref{fig:conv},  we estimate that the numerical error is $\lesssim1\%$, as claimed in the main text.

\end{appendix}

\bibliographystyle{unsrt}
\addcontentsline{toc}{section}{References}
\bibliography{stokesbib}

\end{document}